\documentclass[a4paper,fleqn,usenatbib]{mnras}
\usepackage[T1]{fontenc}
\usepackage{ae,aecompl}

\usepackage{graphicx}	
\usepackage{amsmath}	
\usepackage{amssymb}	
\usepackage{booktabs}
\usepackage{multirow,multicol}
\usepackage{longtable,lscape}
\usepackage{rotating}

\newcommand{\hii}{\hbox{H\,{\sc ii}}}
\newcommand{\oip}{\hbox{O\,{\sc i}\,$\lambda1304$}}
\newcommand{\oi}{\hbox{[O\,{\sc i}]$\lambda6300$}}   
\newcommand{\oiid}{\hbox{[O\,{\sc ii}]$\lambda\lambda3726,3729$}}
\newcommand{\oiit}{\hbox{[O\,{\sc ii}]$\lambda3727$}}
\newcommand{\oiii}{\hbox{[O\,{\sc iii}]$\lambda5007$}}
\newcommand{\oiiinuv}{\hbox{[O\,{\sc iii}]$\lambda2321$}}
\newcommand{\oiiisf}{\hbox{O\,{\sc iii}]}}
\newcommand{\oiiid}{\hbox{O\,{\sc iii}]$\lambda\lambda1661,1666$}}
\newcommand{\oiiit}{\hbox{O\,{\sc iii}]$\lambda1663$}}
\newcommand{\nii}{\hbox{[N\,{\sc ii}]$\lambda6584$}} 
\newcommand{\siid}{\hbox{[S\,{\sc ii}]$\lambda\lambda6717,6731$}}
\newcommand{\siit}{\hbox{[S\,{\sc ii}]$\lambda6724$}}
\newcommand{\siip}{\hbox{S\,{\sc ii}\,$\lambda\lambda1304,1309$}}
\newcommand{\silivoiv}{\hbox{S\,{\sc iv}\,$\lambda1397+$O\,{\sc iv}]$\lambda1402$}}
\newcommand{\silivoivt}{\hbox{S\,{\sc iv}+O\,{\sc iv}]$\lambda1400$}}
\newcommand{\niii}{\hbox{N\,{\sc iii}]$\lambda1750$}}
\newcommand{\nivd}{\hbox{[N\,{\sc iv}]$\lambda1483$+N\,{\sc iv}]$\lambda1487$}}
\newcommand{\nivt}{\hbox{N\,{\sc iv}]$\lambda1485$}}
\newcommand{\nv}{\hbox{N\,{\sc v}\,$\lambda1240$}}
\newcommand{\civd}{\hbox{C\,{\sc iv}\,$\lambda\lambda1548,1551$}}
\newcommand{\civt}{\hbox{C\,{\sc iv}\,$\lambda1550$}}
\newcommand{\ciiid}{\hbox{[C\,{\sc iii}]$\lambda1907$+C\,{\sc iii}]$\lambda1909$}}
\newcommand{\ciiit}{\hbox{C\,{\sc iii}]$\lambda1908$}}
\newcommand{\cii}{\hbox{C\,{\sc ii}]$\lambda2326$}}
\newcommand{\heii}{\hbox{He\,{\sc ii}\,$\lambda1640$}}
\newcommand{\silii}{\hbox{Si\,{\sc ii}\,$\lambda1814$}}
\newcommand{\siliii}{\hbox{Si\,{\sc iii}]}}
\newcommand{\siliiid}{\hbox{[Si\,{\sc iii}]$\lambda1883$+Si\,{\sc iii}]$\lambda1892$}}
\newcommand{\siliiit}{\hbox{Si\,{\sc iii}]$\lambda1888$}}
\newcommand{\neiii}{\hbox{[Ne\,{\sc iii}]$\lambda3343$}}
\newcommand{\neiv}{\hbox{[Ne\,{\sc iv}]$\lambda2424$}}
\newcommand{\nev}{\hbox{[Ne\,{\sc v}]$\lambda3426$}}
\newcommand{\nevbis}{\hbox{[Ne\,{\sc v}]$\lambda3347$}}

\def\apj{\mbox{ApJ}}
\def\apjl{\mbox{ApJL}}
\def\apjs{\mbox{ApJS}}
\def\aaps{\mbox{A\&AS}}
\def\mnras{\mbox{MNRAS}}
\def\aj{\mbox{AJ}}
\def\araa{\mbox{ARA\&A}}
\def\pasp{\mbox{PASP}}

\def\aap{\mbox{A\&A}}
\def\jcap{\mbox{JCAP}}

\def\apss{\mbox{Ap\&SS}}
\def\pasj{\mbox{PASJ}}

\title[Spectral diagnostics of AGN versus star formation]{Nuclear activity versus star formation: emission-line diagnostics at ultraviolet and optical wavelengths}

\author[A. Feltre et al.]{
A. Feltre,$^{1}$\thanks{E-mail: feltre@iap.fr}
S. Charlot,$^{1}$
J. Gutkin$^{1}$
\\
$^{1}$Sorbonne Universit\'es, UPMC-CNRS, UMR7095, Institut d'Astrophysique de Paris, F-75014, Paris, France
}

\pubyear{2015}

\begin{document}
\label{firstpage}
\pagerange{\pageref{firstpage}--\pageref{lastpage}}
\maketitle

\begin{abstract}

In the context of observations of the rest-frame ultraviolet and optical emission from distant galaxies, we explore the emission-line properties of photoionization models of active and inactive galaxies. Our aim is to identify new line-ratio diagnostics to discriminate between gas photoionization by active galactic nuclei (AGN) and star formation. We use a standard photoionization code to compute the emission from AGN narrow-line regions and compare this with calculations of the nebular emission from star-forming galaxies achieved using the same code. We confirm the appropriateness of widely used optical spectral diagnostics of nuclear activity versus star formation and explore new diagnostics at ultraviolet wavelengths. We find that combinations of a collisionally excited metal line or line multiplet, such as \civd, \oiiid, \niii, \siliiid\ and \ciiid, with the \heii\ recombination line are individually good discriminants of the nature of the ionizing source. Diagrams involving at least 3 of these lines allow an even more stringent distinction between active and inactive galaxies, as well as valuable constraints on interstellar gas parameters and the shape of the ionizing radiation. Several line ratios involving Ne-based emission lines, such as \neiv, \neiii\ and \nev, are also good diagnostics of nuclear activity. Our results provide a comprehensive framework to identify the sources of photoionization and physical conditions of the ionized gas from the ultraviolet and optical nebular emission from galaxies. This will be particularly useful to interpret observations of high-redshift galaxies with future facilities, such as the \textit{James Webb Space Telescope} and extremely large ground-based telescopes. 

\end{abstract}

\begin{keywords}
galaxies: active -- galaxies: starburst -- galaxies: high-redshift -- galaxies: quasars: emission lines -- ultraviolet: galaxies
\end{keywords}

\section{Introduction}

Understanding the physical processes responsible for the reionization of the Universe at the end of the Dark Ages is one of the main current problems in astrophysics. According to recent results, this relatively rapid process occurred over the redshift range $6\lesssim z\lesssim 10$ \citep{planck15,robertson15} until the Universe was fully ionized at $z\sim6$ \citep[e.g.][]{fan06}. Star-forming galaxies and active galactic nuclei (AGN) are thought to be main drivers of cosmic reionization, but their relative roles in contributing to the ionizing radiation is still scarcely known \citep[e.g.][]{haardt15}.  The observed drop in the number density of bright quasars at redshifts $z\gtrsim3$ suggests an insufficient contribution by AGN to the observed ionization of the intergalactic medium at high redshift. As a consequence, early star-forming galaxies have been favoured as the main drivers of cosmic reionization \citep[e.g.][]{cowie09,willott10,fontanot12}, although quasars, mini-quasars and a faint AGN population could also play a significant role \citep[e.g.][]{volonteri09,fiore12,haardt15,hao15}. 

Spectroscopic studies of primeval galaxies are a powerful means of probing the nature of the ionizing sources within them. Today, studies of the redshifted ultraviolet and optical emission of distant galaxies by means of near-infrared spectroscopy are confined to small samples of star-forming galaxies \citep[e.g.][]{erb10,stark14} and AGN \citep[e.g.][]{hainline11,steidel12} out to $z\sim3$. Pioneering detections of rest-frame ultraviolet and optical emission lines at higher redshift, out to $z\gtrsim6$, are also mainly limited to few gravitationally lensed sources \citep[e.g.][]{fosbury03,bayliss14,stark15a,stark15b,zitrin15}. In the near future, new facilities, such as the \textit{James Webb Space Telescope} (\textit{JWST}) and extremely large ground-based telescopes (ELT), will collect high-quality, rest-frame ultraviolet and optical spectra of thousands of high-redshift galaxies, allowing extensive investigations of early galaxy evolution and of the relation between star formation and black-hole growth. In fact, spectroscopic analyses of emission-line strengths using photoionization models have proven to be an optimal way of interpreting the signatures of different ionizing sources in galaxies. Studies of the nebular emission from \hii\ regions and AGN often rely on the predictions of standard photoionization codes, such as {\scshape MAPPINGS} \citep{sutherland93,allen08,dopita13} and {\scshape CLOUDY} \citep{ferland93,ferland98,ferland13}, which have both undergone regular updates over the past decades. These studies have shown that important constraints on the density, ionization and metallicity of the photoionized gas can be derived from the relative intensities of ultraviolet and (mainly) optical emission lines. This has led to the identification of standard optical line-ratio diagnostic diagrams to separate nebular emission from active and inactive galaxies, such as the \citet[][hereafter BPT]{baldwin81} and \citet{veilleux87} diagrams based  on the H$\alpha$, H$\beta$, \oiid\ (hereafter simply \oiit), \oiii, \oi, \nii\ and \siid\ (hereafter simply \siit) emission lines. In addition, ultraviolet emission lines have proven to be useful to distinguish between photoionization by AGN and shocks \citep[e.g.][]{villarmartin97,allen98}, while they have not been systematically investigated yet to discriminate between nuclear activity and star formation.

In this paper, we compute a large grid of photoionization models of AGN narrow-line regions, which we combine with recent predictions of the nebular emission from star-forming galaxies (Gutkin et al., in preparation; hereafter G15), to explore new line-ratio diagnostics of photoionization by nuclear activity and star formation in galaxies. To perform a meaningful analysis, we compute photoionization models of AGN narrow-line regions using the same recent version of the {\scshape CLOUDY} code \citep{ferland13} as G15 and adopt their parametrization of the metal and dust content in the ionized gas. We establish the suitability of these models by showing that they confirm previous widely used observational criteria to separate active from inactive galaxies in optical line-ratio diagrams \citep[BPT;][]{veilleux87}. On these grounds, we investigate the extent to which the relative intensities of ultraviolet lines commonly detected in galaxy spectra can allow one to discriminate between photoionization by an AGN and star formation. This leads us to propose new ultraviolet spectral diagnostics of the nature of ionizing radiation, which should be particularly useful to prepare the exploration of galaxies near the reionization epoch with future facilities, such as \textit{JWST} and ground-based ELT.

The paper is structured as follows. In Section~\ref{sec:models}, we describe our models of the nebular emission from narrow-line regions of AGN, as well as the difference between these models and the G15 models of nebular emission from star-forming galaxies. We compare the distributions of the two grids of models in standard optical line-ratio diagnostic diagrams in Section~\ref{sec:optical}. We extend this comparison to different types of ultraviolet line-ratio diagrams in Section~\ref{sec:UV}, with the aim to identify the most promising diagnostics of photoionization by nuclear activity versus star formation in this wavelength range. We summarise our conclusions in Section~\ref{sec:conclusions}, where we also discuss the usefulness of our models to interpret current and future observations of galaxies out to the reionization epoch.

\section{Photoionization models}\label{sec:models}

In this section, we present our models of nebular emission from AGN narrow-line regions (hereafter simply `AGN models'; Section~\ref{sec:AGNNLR}). We also outline the main difference of these models with those of nebular emission from star-forming galaxies by G15 (hereafter simply `SF models'; Section~\ref{sec:G15}). In both cases, nebular emission is computed using the latest version of the photoionization code {\scshape CLOUDY} (c13.03), last described in \cite{ferland13}, in which the gas is assumed to be distributed in concentric spherical layers around the ionizing source (AGN or star cluster).  G15 adopt the approach of \citet[hereafter CL01]{charlot01} to parametrize their models in terms of the characteristic mass of ionizing star clusters and the density, metallicity and dust-to-heavy element mass ratio of photoinized gas. In this approach, the rate of ionizing photons produced by a star cluster evolves as the stellar population ages, and the nebular emission from an entire galaxy is computed by convolving that of individual star clusters with the star formation history (see CL01 for detail). 

In the case of photoionization by an AGN power source (i.e. the accretion disc surrounding a central black hole; see Section~\ref{sec:AGNNLR}), we may assume that the ionizing source does not evolve in time. In this case, the analog of equation~(8) of CL01 for the effective rate of ionizing photons seen by gas irradiated by an accretion disc of luminosity $L_{\rm AGN}$ is
\begin{equation}
Q=L_{\rm AGN}\int_{\nu_{\rm L}}^{\infty}{d\nu}\frac{S_{\nu}}{h\nu}\,,
\label{eq:q}
\end{equation}
where $h$ and $c$ are, respectively, the Planck constant and the speed of light and $\nu_{\rm L}$ is the frequency at the Lyman limit, $h\nu_{\rm L}=13.6\,$eV. In this expression, the quantity $S_{\nu}$ is the luminosity per unit frequency of the accretion disc (equation~\ref{eq:Lagn} below). The ionization parameter at the distance $r$ from the ionizing source, $U(r)$, is defined as the dimensionless ratio of the number density of H-ionizing photons to that of hydrogen, $n_{\rm H}$, i.e.
\begin{equation}
U(r)={Q}\slash({4\pi r^2 n_{\rm H} c})\,,
\label{eq:u}
\end{equation}
where $Q$ is the rate of the ionizing photons (in ${\rm s}^{-1}$). As recalled by CL01, the actual geometry of a model and the ionization profile of the gas are set by the ratio between the innermost radius and the Str\"omgren radius, $R_{\rm S}$, defined as 
\begin{equation}
R_{\rm S}^3={3Q}\slash{(4\pi n_{\rm H}^2 \epsilon \alpha_{\rm B})}\,, 
\label{eq:rs}
\end{equation}
where $\epsilon$ the volume-filling factor of the gas (i.e. the ratio of the volume-averaged hydrogen density to $n_{\rm H}$) and $\alpha_{\rm B}$ the case-B hydrogen recombination coefficient \citep{osterbrock06}. The geometry is truly spherical when the inner radius of the gaseous nebula is such that $r_{\rm in}\ll R_{\rm S}$, i.e. the thickness of the ionized region reaches the order of magnitude of $R_{\rm S}$, while in the case $r_{\rm in} \gtrsim R_{\rm S}$, the geometry is plane-parallel. We parametrize our models below in terms of the ionization parameter at the Str\"omgren radius, given by (equations~\ref{eq:u}--\ref{eq:rs})
\begin{equation}
U_{\rm S} =\frac{\alpha_{\rm B}^{2/3}}{3c} \left( \frac{3 Q\epsilon^{2} n_{\rm H}}{4\pi} \right)^{1/3}\approx\langle U \rangle/3\,,
\label{eq:us}
\end{equation}
where $\langle U \rangle$ is the volume-averaged ionization parameter and the near-equality arises from our neglect of the weak dependence of $\alpha_{\rm B}$ on $r$ through the electron temperature.

All calculations presented in this paper account for the depletion of metals onto dust grains in the photoionized gas and the associated absorption and scattering of incident radiation, radiation pressure, collisional cooling and photoelectric heating of the gas (see, e.g., \citealt{ferland96}, but also \citealt{shields95}, \citealt{dopita02} and \citealt{groves04a} for a description of these effects on the emergent nebular emission). One of the most notable effects of the depletion of metals from the gas phase is the less efficient cooling through infrared fine-structure transitions, which increases the electron temperature, and hence, cooling through the more energetic optical transitions. We adopt the default dust properties of {\scshape CLOUDY}, consisting of a mixture of graphite and silicates with a standard \citet{MRN} grain size distribution and optical properties from \cite{martin91}. In the following subsections, we describe in more detail the parametrization of our photoionization models of AGN and star-forming galaxies.

\subsection{Narrow-line regions of AGN}\label{sec:AGNNLR}

At least 3 approaches have been adopted in the past to parametrize the gas distribution in photoionization models of AGN narrow-line regions: (i) a combination of matter- and ionization-bounded clouds \citep{binette96}; (ii) the `locally emitting clouds' model \citep{ferguson97}, which combines clouds of different densities at different distances from the ionizing source; and, more recently, (iii) clouds of a single type, but including dust and dominated by radiation pressure \citep{dopita02,groves04a}. This third kind of approach has been shown to provide good agreement with observations over a wide range of ionization parameters (most recently by, e.g., \citealt{richardson14}). It is therefore the one we adopt in the present work.

We adopt the so-called `open geometry' in {\scshape CLOUDY}, appropriate for gas with small covering factor \citep[typically less than 10 per cent for AGN narrow-line regions;][]{maiolino01,baskin05}, and describe the physical conditions of the photoionized gas in terms of the following main adjustable parameters, also listed in Table~\ref{tab:parameters}.

\begin{itemize}

\item {\it Luminosity per unit frequency $S_{\nu}$ of the accretion disc.} The emission from the accretion disc in an AGN is usually approximated by a broken power law, for which we adopt here the form
\begin{equation}
S_\nu \propto
\left \{
\begin{array}{ll}
\nu^{\alpha} & \mbox{at wavelengths} \quad  0.001\leq  \lambda/\micron \leq 0.25\,,\\
\nu^{-0.5} & \mbox{at wavelengths} \quad  0.25< \lambda/\micron \leq 10.0\,,\\ 
\nu^{2} & \mbox{at wavelengths} \quad  \lambda/\micron >10.0\,.\\
\end {array}
\right.
\label{eq:Lagn}
\end{equation}
Hence, we keep the power-law index $\alpha$ at ultraviolet and optical wavelengths as an adjustable parameter.  We explore values of this parameter in the range $-2.0\leq\alpha\leq-1.2$, encompassing values often adopted in the literature \citep[e.g.,][]{groves04a} and consistent with observational constraints. For example, \cite{zheng97} find $\alpha\approx-1.8$ in the composite spectrum of 41 radio-quiet quasars at redshifts $0.3\lesssim z\lesssim1.5$ observed with the  \textit{HST} Faint Object Spectrograph, while \cite{lusso15} find the stacked ultraviolet spectrum of 53 quasars at $z\sim2.4$ observed with the \textit{HST} Wide Field Camera 3 to have a slope of $\alpha\approx-1.7$ blueward of Ly$\alpha$. The power-law indices at longer wavelengths in expression~(\ref{eq:Lagn}) are taken from \cite{feltre12}. We adopt a fixed accretion-disc luminosity $L_{\rm AGN}=10^{45}$~erg~s$^{-1}$~cm$^{-2}$ and an inner radius of the narrow-line region $r_{\rm in}\approx300$~pc, corresponding to an incident flux of $L_{\rm AGN}/4\pi r_{\rm in}^2 \approx 10^2$~erg~s$^{-1}$~cm$^{-2}$ \citep{netzer13}.

\item {\it Ionization parameter $U_{\rm S}$ at the Str\"omgren radius}. This is related to the gas density, $n_{\rm H}$, the volume-filling factor of the gas, $\epsilon$, and the rate of ionizing photons, $Q$, via equation~(\ref{eq:us}). We consider below values of this parameter in the range $-4.0\leq\log U_{\rm S}\leq-1.0$. We note that other authors sometimes use the ionization parameter  at the inner edge $r_{\rm in}$ of the narrow-line region, noted $U_{0}$, as an adjustable parameter \citep[e.g.][]{groves04b,nagao06}. According to equation~(\ref{eq:u}), $U_{0}$ will be larger than $U_{\rm S}$ for $r_{\rm in}\ll R_{\rm S}$ and more similar to $U_{\rm S}$ for $r_{\rm in}\gtrsim R_{\rm S}$.

\item {\it Hydrogen number density $n_{\rm H}$}. We consider values of this parameter in the range $10^2\leq n_{\rm H}/{\rm cm}^3\leq 10^{4}$ (we also run test models with densities up to $n_{\rm H}=10^{6}$~cm$^{-3}$).

\item {\it Gas metallicity $Z$}. We adopt the same relative abundances of heavy elements as in G15, taken from \citet[][largely based on \citealt{caffau11}]{bressan12} for solar metallicity. All heavy elements except nitrogen are assumed to scale linearly with oxygen abundance. Secondary production of nitrogen is accounted for following the prescription of \citet{groves04a},
\begin{equation}\label{eq:Ngroves04}
{\rm [N/H]}={\rm [O/H]}\left(10^{-1.6}+ 10^{(2.33+\log_{10}{\rm [O/H]})}\right)\,.
\end{equation}
The nitrogen abundance is properly rescaled to account for the difference in solar abundance between \cite{bressan12} and \cite{groves04a}, as outlined in G15. The helium abundance is given by \citep{bressan12}
\begin{equation}\label{eq:He}
Y=Y_{\rm P} + (Y^{0}_{\odot} - Y_{\rm P}) Z/Z^{0}_{\odot} = 0.2485 + 1.7756 Z\,,
\end{equation}
where $Y_{\rm P}=0.2485$ \citep[][see also \citealt{steigman10}]{komatsu11} and $Y_{\odot}^{0}=0.28$ are, respectively, the primordial and  zero-age main sequence solar (i.e. before diffusion; see \citealt{bressan12}) abundances of helium, and $Z^{0}_{\odot}= 0.01774$ is the  zero-age main sequence solar metallicity. The present-age solar metallicity adopted in G15 and in this work is $Z_{\odot}= 0.01524$ \citep{bressan12}. In what follows, we compute AGN models for 15 values of the metallicity across the range $0.0001\leq Z\leq0.07$.

\item {\it Dust-to-heavy element mass ratio $\xi_{\rm d}$}. This accounts for the depletion of metals onto dust grains in the ionized gas. To explore the impact of the presence of dust on the emission line properties and diagnostics, following CL01 and G15, we consider values of the dust-to-metal mass ratio in the range $0.1\leq\xi_{\rm d}\leq0.5$.

\end{itemize}

\begin{table*}
\begin{center}
\begin{tabular}{ccc}
\hline
 Parameter & AGN narrow-line regions & Star-forming galaxies \\
\hline
\hline
Ionizing spectrum & $\alpha= -1.2, -1.4, -1.7, -2.0$ & constant star formation rate, age = $10^{8}\,$yr \\
\hline
$\log U_{\rm S}$ & $-1.0, -2.0,  -3.0 , -4.0$ &  $-1.0, -1.5, -2.0, -2.5, -3.0, -3.5, -4.0$\\
\hline
$\log(n_{\rm H}/{\rm cm}^{-3})$ & 2.0, 3.0, 4.0 & 2.0, 3.0, 4.0\\  
\hline
$Z$ & 0.0001, 0.0002, 0.0005, 0.001, 0.002, 0.004, 0.006, & 0.0001, 0.0002, 0.0005, 0.001, 0.002, 0.004, 0.006,\\ & 0.008, 0.014, 0.01774, 0.03, 0.04, 0.05, 0.06, 0.07 & 0.008, 0.014, 0.01774, 0.03\\
\hline
$\xi_{\rm d}$ & 0.1, 0.3, 0.5 & 0.1, 0.3, 0.5\\
\hline
\end{tabular}
\caption{Adjustable parameters of the AGN narrow-line region (second column) and SF (third column) photoionization models. See Section~\ref{sec:models} for details.}
\end{center}
\label{tab:parameters}
\end{table*}

\subsection{Star-forming galaxies}\label{sec:G15}

The main motivation of the present work is to identify emission-line diagnostics of active versus inactive galaxies. To this end, we compare the AGN photoionization models described in the previous subsection with the models of the nebular emission from star-forming galaxies computed by G15. These calculations combine the latest version of the \citet{bruzual03} stellar population synthesis model (Charlot \& Bruzual, in preparation) with the photoionization code {\scshape CLOUDY}, following the approach outlined by CL01. This approach consists in convolving the spectral evolution of single, ionization bounded \hii\ regions with a star formation history to compute the nebular emission of a whole galaxy. In this context, the parameters of the photoionization model should be interpreted as effective (i.e. galaxy-wide) ones, describing the ensemble of \hii\ regions and the diffuse gas ionized by stars throughout the galaxy. A `closed geometry' is used in {\scshape CLOUDY} to perform these calculations, as appropriate for spherical \hii\ regions.

G15 ran a large grid of SF models encompassing models with both solar and non-solar C/O abundance ratio and different stellar initial mass functions and star formation histories. We consider here their models for a standard, solar C/O abundance ratio of 0.44, a standard  \cite{chabrier03} initial mass function truncated at 0.1 and $100\,M_\odot$ and constant star formation rate, at an age of 100\,Myr. We note that the star formation history and age have a negligible influence on the predicted emission-line ratios, so long as star formation has been constant for at least 10\,Myr, corresponding to the age over which 99.9 per cent of the ionizing photons are released by a single stellar generation (e.g. CL01). In the rightmost column of Table~\ref{tab:parameters}, we list the ranges in ionization parameter, $-4.0\leq\log U_{\rm S}\leq-1.0$, gas density, $10^2\leq n_{\rm H}/{\rm cm}^3\leq 10^4$, metallicity, $0.0001\leq Z\leq0.03$, and dust-to-metal mass ratio, $0.1\leq\xi_{\rm d}\leq0.5$, covered by the grid of G15 models considered here. In these calculations, the metallicity of the ionizing stars is always taken to be the same as that of the photoionized gas.

\begin{figure*}
\begin{center}
\includegraphics[width=10.0cm, angle=270]{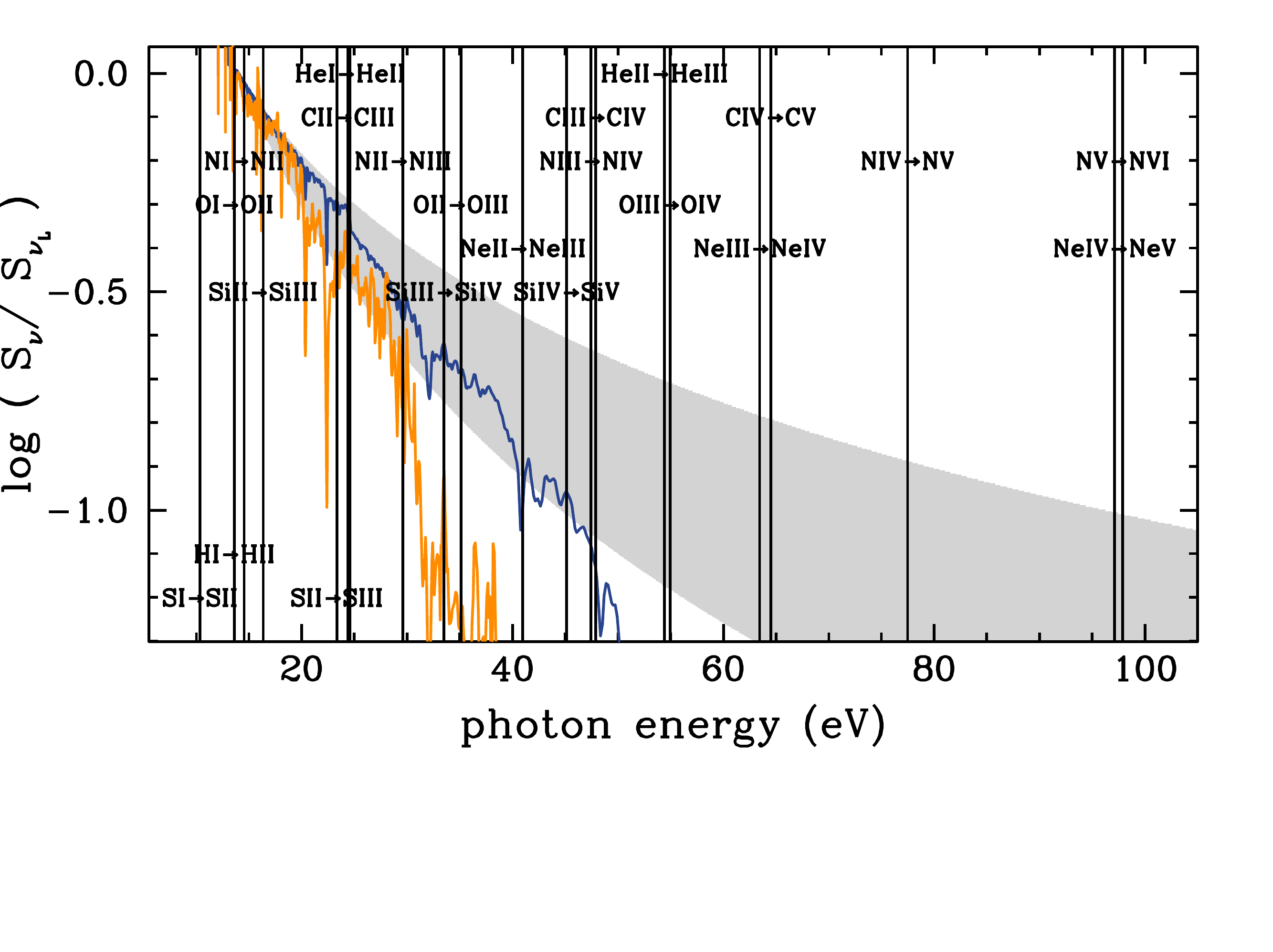} 
\caption{Examples of spectral energy distributions (in units of the luminosity per unit frequency at the Lyman limit) of the incident ionizing radiation in the AGN and SF models of Table~\ref{tab:parameters}. The grey shaded area indicates the location of AGN ionizing spectra with power-law indices between $\alpha=-2.0$ (bottom edge) and $-1.2$ (top edge). The blue and orange lines show the ionizing spectra of stellar populations of metallicities $Z=0.001$ and 0.03, respectively. Vertical lines indicate the ionizing energies of ions of different species.}
\label{fig:ion_spec}
\end{center}
\end{figure*}

The main feature differentiating models of active versus inactive galaxies in Table~\ref{tab:parameters} is the spectral energy distribution of the incident ionizing radiation. In this respect, a comparison between the input ionizing spectra associated with AGN and SF models is instructive to understand the differences in the predictions of these models outlined in Sections~\ref{sec:optical} and \ref{sec:UV}. In Fig. \ref{fig:ion_spec} (inspired from fig.~19 of \citealt{steidel14} and fig.~4 of \citealt{stark15b}), we compare the spectral energy distributions of AGN models with various spectral indices (grey shaded area) to that of SF models with two metallicities, $Z=0.001$ (blue line) and 0.03 (orange line). The two stellar population spectra start to differ significantly at photon energies greater than $20\,$eV, the lowest-metallicity one being the hardest, as low-metallicity stars evolve at higher temperatures and luminosities than high-metallicity ones \citep[e.g.,][]{chen15}. We also note that up to energies of $\sim30\,$eV, the spectra of AGN accretion discs and star-forming galaxies are relatively similar, while above $\sim50\,$eV, the stellar population spectra drop abruptly. Thus, emission lines associated with ion species requiring ionization energies greater than $50\,$eV should be the most promising indicators of the presence of an AGN, as we shall see in the following sections \citep[see also][]{steidel14,stark15b}.

\section{Optical emission lines and standard AGN/star-formation diagnostics}\label{sec:optical}

The luminosity ratios of strong optical emission lines are standard diagnostics of the source of the ionizing radiation in external galaxies, as they allow one to discriminate between the spectra of \hii\ regions (i.e. star-forming galaxy) and those of type-2 AGN \citep[i.e. narrow-line regions; e.g. BPT;][]{veilleux87}. As a result, diagnostic diagrams defined by optical lines ratios, such as \oiii/H$\beta$, \nii/H$\alpha$, \siit/H$\alpha$ and \oi/H$\alpha$, are widely used in the literature to interpret observations of the nebular emission from active and inactive galaxies \citep[e.g.][]{dopita02,kauffmann03,groves04b,kewley01,kewley06,kewley13a,kewley13b}. In this section, we show that the models presented in Section~\ref{sec:models} reproduce well observations of nearby galaxies powered by either star formation or an AGN in different standard optical line-ratio diagrams. We also investigate the dependence of these line ratios on the adjustable parameters of the models.

\subsection{SDSS observational sample}\label{sec:SDSS}

To establish the usefulness of our AGN and SF models to interpret observations of the nebular emission from galaxies, we appeal to high-quality observations of nearby galaxies from the Sloan Digital Sky Survey Data Release 7 \citep[SDSS DR7;][]{abazajian09}.\footnote{We use the emission-line measurements made available online by the Max-Planck-Institute for Astrophysics (Garching) and Johns Hopkins University (MPA-JHU) collaboration in the file \texttt{SepcObjAll.fits} at http://home.strw.leidenuniv.nl/$\sim$jarle/SDSS/.} By analogy with \citet[][see also \citealt{kewley06,yuan10}]{juneau14}, we select all primary targets (SCIENCEPRIMARY = 1) in the redshift range $0.04\leq z \leq0.2$. The lower redshift cut limits the influence of strong aperture effects, while the upper one allows the detection of galaxies with intrinsically weak emission lines while increasing the statistics on Seyfert~2 galaxies \citep{juneau14}. We consider only those galaxies with measurements available simultaneously for all six optical emission lines entering the definitions of the diagnostics mentioned above, i.e., H$\beta$, \oiii, \oi, H$\alpha$, \nii\ and \siit. Following \citet{juneau14}, we further select galaxies according to signal-to-noise ratio (S/N) in line-ratio rather than individual-line measurements, which allows the sampling of a wider range of intrinsic emission-line properties (we use the factors in table~4 of \citealt{juneau14} to scale up the formal errors in flux-ratio uncertainties in the MPA-JHU catalogs, as inferred from the results of duplicate observations of the same galaxies). We adopt the same criterion as these authors and require ${\rm S/N}>2.12$ ($=3/\sqrt{2}$) in the emission-line ratios of interest to us, i.e., \oiii/H$\beta$, \nii/H$\alpha$, \siit/H$\alpha$ and \oi/H$\alpha$. This leaves us with a final sample of 202,083 galaxies with high-quality line-ratio measurements. We correct the emission-line ratios for attenuation by dust, based on the departure of the observed Balmer decrement from the dust-free case-B recombination value and using the standard reddening curve of \cite{cardelli89}.

\subsection{\oiii/H$\beta$ versus \nii/H$\alpha$ diagram}\label{sec:BPT}

\begin{figure*}
\begin{center}
\includegraphics[width=17.0cm]{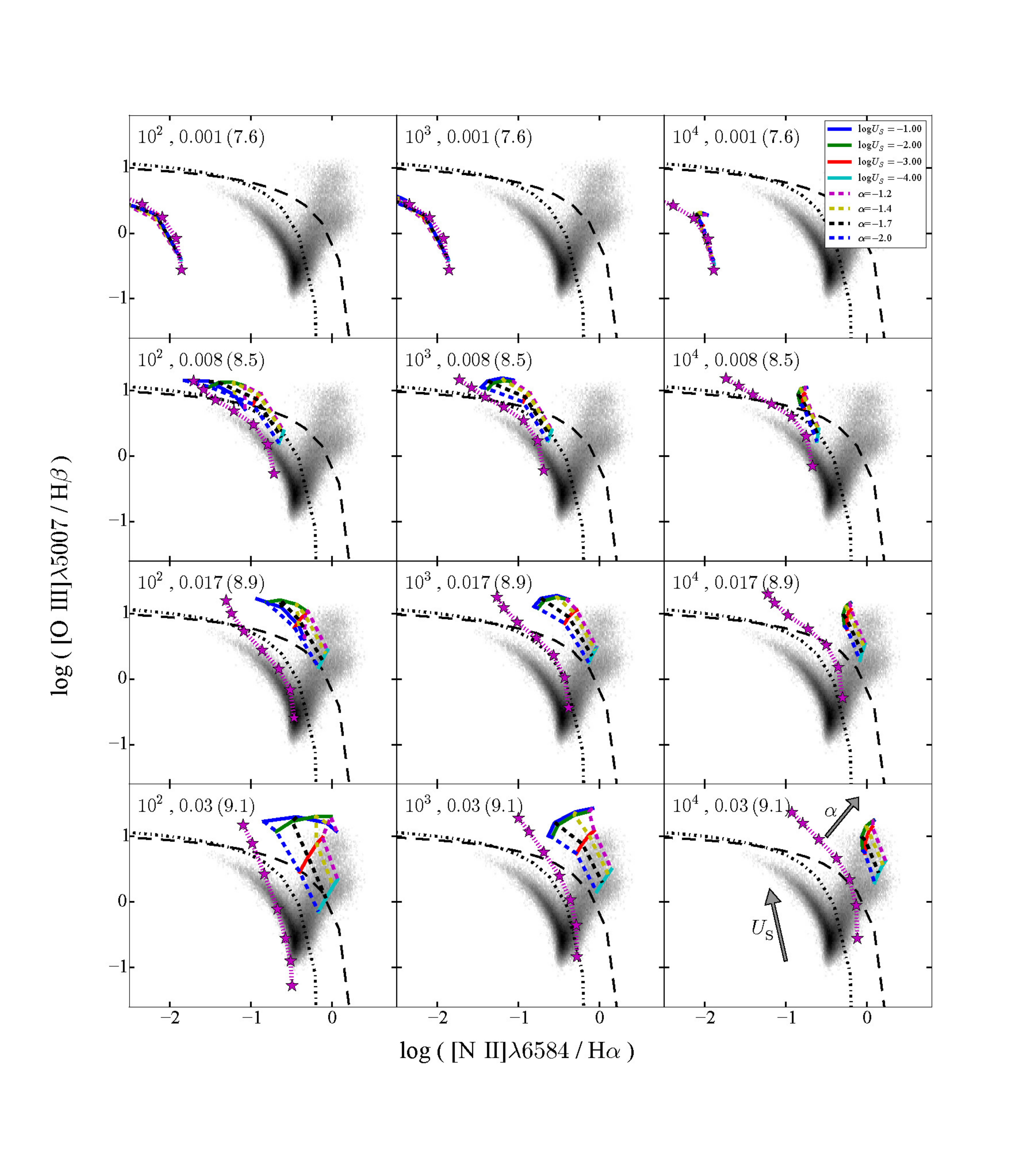} 
\caption{Predictions of the AGN and SF models described in Section~\ref{sec:models} in the standard \oiii/H$\beta$ versus \nii/H$\alpha$ BPT diagnostic diagram, for different assumptions about the gas density ($n_{\rm H}=10^2$, $10^3$ and $10^4$~cm$^{-3}$ from left to right) and metallicity ($Z=0.0001$, 0.001, 0.01774, and 0.03 from top to bottom). Also indicated in parenthesis in each panel is the interstellar oxygen abundance, $12+\log{\rm (O/H)}$, which includes the components in the gas and dust phases. In each panel, we show AGN models corresponding to wide ranges in power-law index, $-2.0\leq\alpha\leq-1.2$, and ionization parameter, $-4.0\leq\log U_{\rm S}\leq-1.0$ (both increasing from bottom to top as indicated in the lower-right panel, and colour-coded as indicated in the top-right panel), and SF models (magenta stars) corresponding to the same range in $U_{\rm S}$. In each panel, the data are the SDSS observations described in Section~\ref{sec:SDSS}, while the long-dashed and dot-dashed black lines indicate, respectively, the criteria of \protect\cite{kewley01} and \protect\cite{kauffmann03} to distinguish AGN from star-forming galaxies.}
\label{fig:BPT1}
\end{center}
\end{figure*}

We start by considering the standard \oiii/H$\beta$ versus \nii/H$\alpha$ diagram originally proposed by BPT to discriminate between AGN and star-forming galaxies. Fig.~\ref{fig:BPT1} shows several versions of this diagram, in which we plot AGN and SF models with different input parameters over the SDSS observations described in Section~\ref{sec:SDSS}. Each panel corresponds to different assumptions about the gas density, $n_{\rm H}$ (increasing from left to right), and metallicity, $Z$ (increasing from top to bottom), as indicated. Also indicated in parenthesis is the interstellar oxygen abundance, which includes the components in the gas and dust phases. In each panel, we show AGN models corresponding to wide ranges in power-law index, $\alpha$, and ionization parameter, $U_{\rm S}$, and SF models corresponding to the same range in $U_{\rm S}$. The long-dashed and dot-dashed black lines indicate, respectively, the popular criteria of \cite{kewley01} and \cite{kauffmann03} to distinguish AGN from star-forming galaxies. 

\begin{figure*}
\begin{center}
\includegraphics[width=15.0cm]{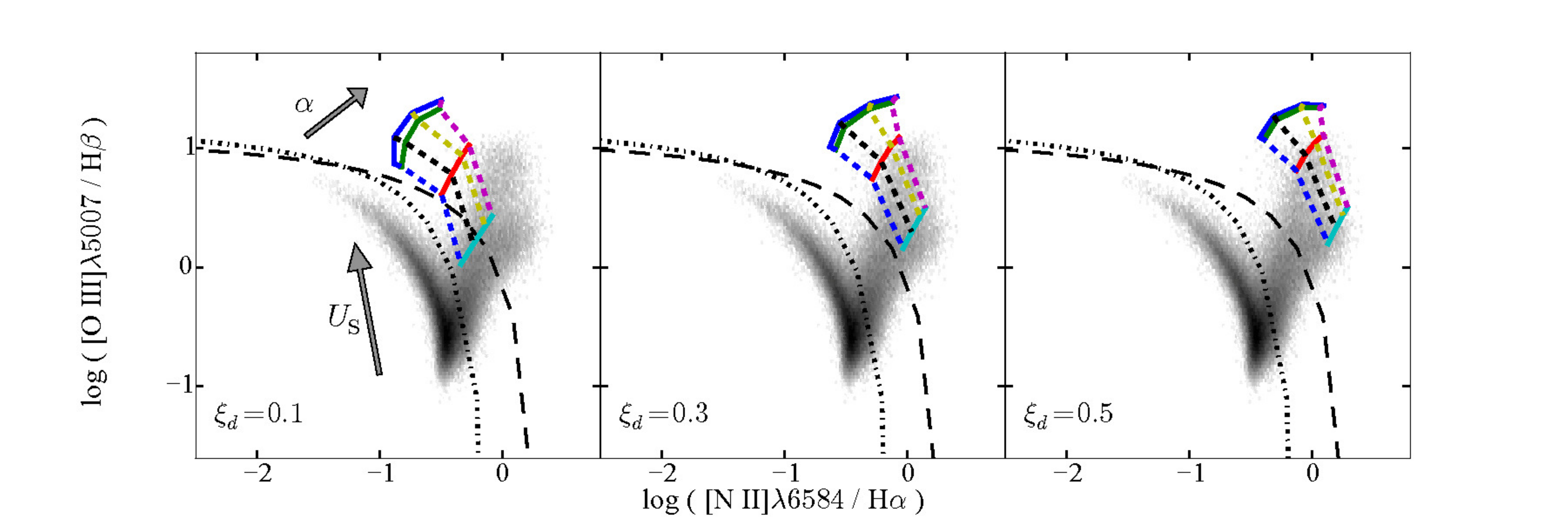} 
\caption{Predictions of the AGN models described in Section~\ref{sec:models} in the \oiii/H$\beta$ versus \nii/H$\alpha$ BPT diagram, for a gas density $n_{\rm H}=10^3$~cm$^{-3}$, a metallicity $Z=0.03$ and dust-to-metal mass ratios $\xi_{\rm d}=0.1$, 0.3 and 0.5, from left to right.  In each panel, we show AGN models corresponding to wide ranges in power-law index, $-2.0\leq\alpha\leq-1.2$, and ionization parameter, $-4.0\leq\log U_{\rm S}\leq-1.0$. The data and colour coding of the models are the same as in Fig.~\ref{fig:BPT1}. The long-dashed and dot-dashed black lines indicate, respectively, the criteria of \protect\cite{kewley01} and \protect\cite{kauffmann03} to distinguish AGN from star-forming galaxies.}
\label{fig:BPT_xid}
\end{center}
\end{figure*}

It is important to note that the parameter ranges considered in Fig.~\ref{fig:BPT1} encompass wider combinations of $n_{\rm H}$, $Z$ and $U_{\rm S}$ than typically observed in the nearby Universe. For example, low-metallicity AGN are extremely rare in local surveys such as the SDSS \citep{groves06}, while in star-forming galaxies, there appears to be an anticorrelation between ionization parameter and metallicity, in the sense that metal-rich galaxies are generally found to have only modest ionization parameters \citep[e.g.,][]{brinchmann04}. Also, gas densities estimated from optical line-doublet analyses are typically around $n_{\rm H}\sim10^3$~cm$^{-3}$ for AGN narrow-line regions and $n_{\rm H}\sim10^2$~cm$^{-3}$ for SF models \citep[e.g., sections 5.6 and 13.4 of][]{osterbrock06}. This should be kept in mind when examining the location of models relative to SDSS data in Fig.~\ref{fig:BPT1}.

Fig.~\ref{fig:BPT1} (top panels) shows that, at low metallicity ($Z\lesssim0.008$), AGN and SF models predict similar \oiii/H$\beta$ and \nii/H$\alpha$ optical line ratios, on the side of the BPT diagram corresponding to star-forming galaxies. This is expected from the similarity of the spectra of accretion discs and low-metallicity stellar populations at the ionizing energies of these ions (Fig.~\ref{fig:ion_spec}). As metallicity increases (from top to bottom in Fig.~\ref{fig:BPT1}), AGN and SF models start occupying distinct regions of the BPT diagram, on either side of the \citealt{kewley01} and \cite{kauffmann03} criteria (SF models falling in the AGN part of the diagram are those combining high $U_{\rm S}$ and high $Z$). This is consistent with the results from previous studies that gas in local AGN is generally found to be metal-rich, with $9.0< \log{\rm (O/H)}+12<9.3$ \citep[e.g.][]{kewley02,groves04a}. The rise in \oiii/H$\beta$ and \nii/H$\alpha$ for both AGN and SF models as metallicity increases in Fig.~\ref{fig:BPT1} follows from the increase in the abundance of coolants. The effect is more pronounced for \nii/H$\alpha$ than for \oiii/H$\beta$ because of secondary nitrogen production at $Z \gtrsim 0.1Z_{\odot}$ \citep[][Section~\ref{sec:AGNNLR} above]{groves04b}. At the highest metallicity, \oiii/H$\beta$ and \nii/H$\alpha$ start to decline again, because the high efficiency of cooling makes the electronic temperature (and hence collisional excitation) drop. We also note in Fig.~\ref{fig:BPT1} that, as expected from Fig.~\ref{fig:ion_spec}, flattening the ionizing spectrum by increasing the power-paw index $\alpha$ makes both \oiii/H$\beta$ and \nii/H$\alpha$ larger for AGN models, the effect being more pronounced at high than at low metallicity and ionization parameter. The SDSS observations are compatible with the entire explored range of spectral slopes.

Fig.~\ref{fig:BPT1} further confirms that, as noted by \cite{groves04b}, the \oiii/H$\beta$ and \nii/H$\alpha$ optical emission-line ratios tend to be less sensitive to changes in density, over the range from $10^2$ to $10^4\, {\rm cm}^{-3}$, than to changes in metallicity and ionization parameter. When $n_{\rm H}$ exceeds the critical density for which the efficiency of collisional de-excitation reaches that of radiative cooling for infrared fine-structure transitions (typically a few times $10^3$~cm$^{-3}$), the electronic temperature rises \citep{stasinska90}. Radiative cooling through optical transitions, which have higher critical densities (about $10^5$~cm$^{-3}$ for \nii\ and $10^6$~cm$^{-3}$ for \oiii), increases, causing \oiiit/H$\beta$ and especially \nii/H$\alpha$ to increase. The effect is more pronounced at high than at low metallicity, and for SF than for AGN models (\oiiit/H$\beta$ even slightly drops for $n_{\rm H}\gtrsim10^4$~cm$^{-3}$ in AGN models with the highest $U_{\rm S}$). We note that, at metallicities $Z\gtrsim0.017$ and gas densities $n_{\rm H}\gtrsim10^3$~cm$^{-3}$, AGN models with ionization parameters across the full explored range ($-4 \leq\log U_{\rm S} \leq-1$) are compatible with SDSS data in Fig.~\ref{fig:BPT1}. At lower $n_{\rm H}$, only ionization parameters $\log U_{\rm S} \lesssim-3$ are favored. For completeness, we also computed AGN models gas densities $n_{\rm H}=10^5$ and $10^6$~cm$^{-3}$ (not shown). While models with $n_{\rm H}=10^5$~cm$^{-3}$ are still marginally consistent with SDSS data in Fig.~\ref{fig:BPT1}, at least for metallicities in the range $0.017\lesssim Z \lesssim 0.03$, this is not the case for those with $n_{\rm H}=10^6$~cm$^{-3}$. The inappropriateness of models with extreme $n_{\rm H}$ to reproduce observations is consistent with the conclusion reached by \cite{nagao06} from the analysis of ultraviolet emission lines in narrow-line regions in high-redshift AGN (using a similar, but simplified model). 

\begin{figure*}
\begin{center}
 \begin{tabular}{ccc}
\includegraphics[width=15.0cm]{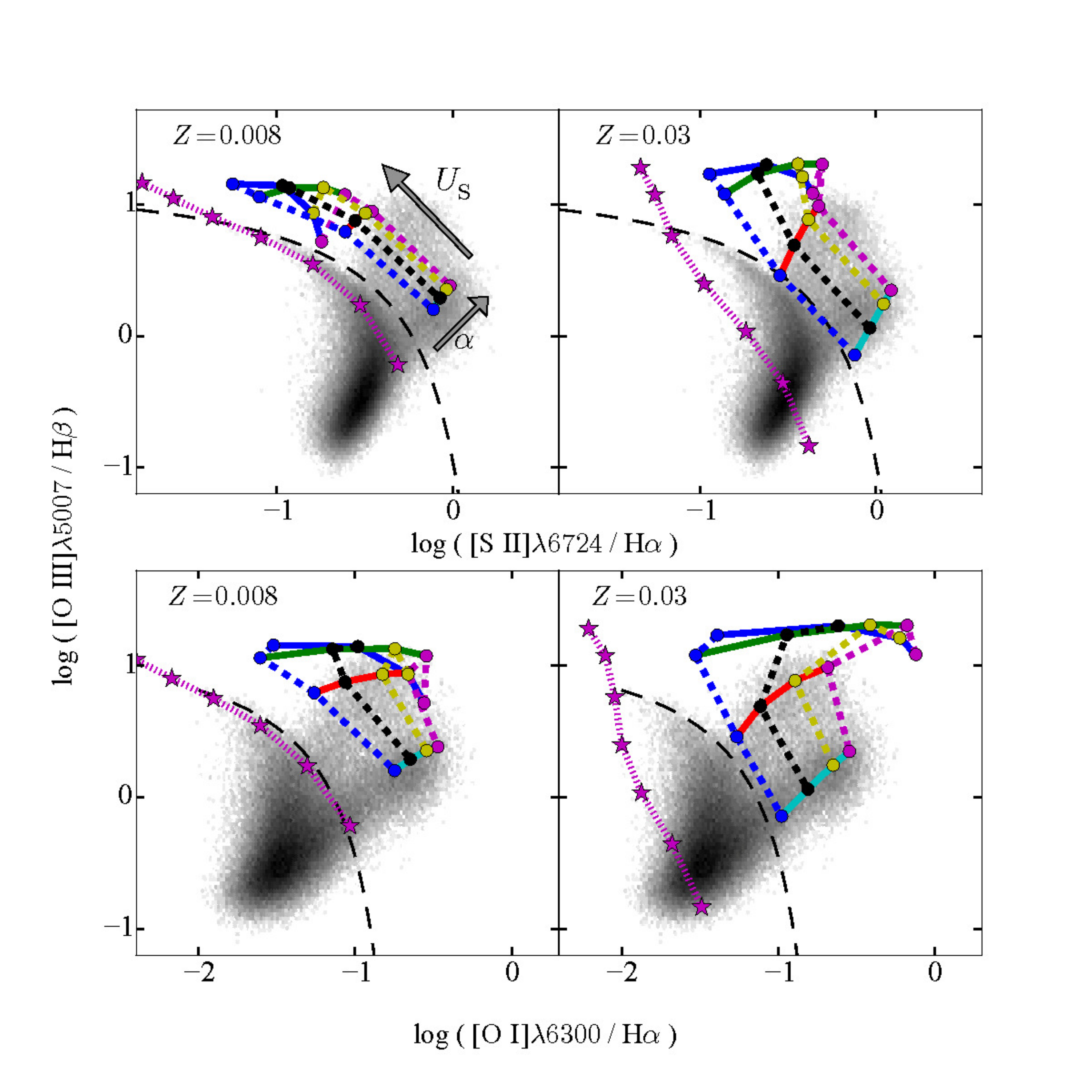} 
 \end{tabular}
\caption{Predictions of the AGN and SF models described in Section~\ref{sec:models} in the \oiii/H$\beta$ versus \siit/H$\alpha$ (top) and \oiii/H$\beta$ versus \oi/H$\alpha$ (bottom) diagrams of \citet{veilleux87}, for two values of the metallicity, $Z=0.008$ (left) and 0.03 (right), and a dust-to-metal mass ratio $\xi_{\rm d}=0.3$. In each panel, we show AGN models corresponding to wide ranges in power-law index, $-2.0\leq\alpha\leq-1.2$, and ionization parameter, $-4.0\leq\log U_{\rm S}\leq-1.0$ (both increasing from bottom to top as indicated in the upper-left panel, and colour-coded as in Fig.~\ref{fig:BPT1}), and SF models corresponding to the same range in $U_{\rm S}$. AGN and SF models have gas densities $n_{\rm H}=10^3$ and $10^2$~cm$^{-3}$, respectively. The long-dashed black line indicates the criterion of \protect\cite{kewley01} to distinguish AGN from star-forming galaxies.}
\label{fig:BPT23}
\end{center}
\end{figure*}

It is worth pointing out that, unlike \cite{dopita02} and \cite{groves04b}, who show that dusty, radiation-pressure dominated models are able to reproduce in a natural way observed ultraviolet and optical emission-line properties of AGN across a wide range of ionization parameters, \cite{nagao06} reach a different conclusion in claiming that high-ionization AGN are best reproduced by dust-free models. Fig.~\ref{fig:BPT_xid} illustrates the effect of metal depletion onto dust grains in our AGN models in the \oiii/H$\beta$ versus \nii/H$\alpha$ BPT diagram. We show models corresponding to $n_{\rm H}=10^3$~cm$^{-3}$, $Z=0.03$ and dust-to-metal mass ratios $\xi_{\rm d}=0.1$, 0.3 and 0.5, from left to right. As $\xi_{\rm d}$ rises, the removal of coolants from the gas phase reduces the cooling efficiency through infrared-fine structure transitions. This makes the electronic temperature rise, and hence cooling through the optical transitions larger \citep{shields95,charlot01}. For important refractory coolants, such as oxygen, the effect of depletion is compensated by the rise in electronic temperature, in such a way that \oiii/H$\beta$ increases only slightly in Fig.~\ref{fig:BPT_xid} as $\xi_{\rm d}$ increases. The emission-lines strengths of non-refractory elements, such as nitrogen and sulfur, increase more significantly, as illustrated by the implied significant rise in \nii/H$\alpha$.

\subsection{Other AGN/star-formation diagnostic diagrams}\label{sec:others}

Other optical line-ratio diagrams have been proposed to discriminate between AGN and star-forming galaxies, such as the \oiii/H$\beta$ versus \siit/H$\alpha$ and \oiii/H$\beta$ versus \oi/H$\alpha$ diagrams \citep{veilleux87}. We show in Fig.~\ref{fig:BPT23} how our AGN and SF models behave in these diagrams, for two values of the metallicity, $Z=0.008$ and 0.03, a dust-to-metal mass ratio in the middle of the explored range, $\xi_{\rm d}=0.3$, and the same ranges in power-law index, $\alpha$, and ionization parameter, $U_{\rm S}$, as in Fig.~\ref{fig:BPT1}. For simplicity, we adopt a single choice of gas density, which differs for the AGN models, $n_{\rm H}=10^3$~cm$^{-3}$, and the SF models, $n_{\rm H}=10^2$~cm$^{-3}$. As mentioned above, these values are those typically estimated from optical line-doublet analyses in nearby active and inactive galaxies \citep[e.g., sections 5.6 and 13.4 of][]{osterbrock06}. 

The results of Fig.~\ref{fig:BPT23} are qualitatively similar to those obtained for the standard \oiii/H$\beta$ versus \nii/H$\alpha$ BPT diagram in Fig.~\ref{fig:BPT1}, confirming the success of our AGN and SF models to account for the observed optical emission-line properties of active and inactive galaxies, on either side of the \cite{kewley01} separation criterion (long-dashed line in Fig.~\ref{fig:BPT23}). In fact, by exploring models (not shown) across the full ranges in $n_{\rm H}$, $Z$ and $\xi_{\rm d}$ considered in Figs~\ref{fig:BPT1} and \ref{fig:BPT23}, we find that the dependence of \siit/H$\alpha$ and \oi/H$\alpha$ on these parameters is similar to that of \oiii/H$\beta$ and \nii/H$\alpha$ described in Section~\ref{sec:BPT}. We conclude that the photoionization models presented in Section~\ref{sec:models} reproduce well the observed optical properties of AGN narrow-line regions and star-forming galaxies in the nearby Universe. These models therefore provide a solid foundation to explore the emission-line properties of active and inactive galaxies at ultraviolet wavelengths.

\section{Ultraviolet emission lines and new AGN/star-formation diagnostics}\label{sec:UV}

In this section, we examine the ultraviolet properties of the photoionization models of AGN narrow-line regions and star-forming galaxies presented in Section~\ref{sec:models}, which we showed in Section~\ref{sec:optical} to reproduce well the optical emission-line properties of nearby active and inactive galaxies. We consider primarily, but not exclusively, far- and near-ultraviolet emission lines commonly detected in the spectra of high-redshift galaxies, with the goal to identify the best spectral diagnostics to discriminate between AGN and star-forming galaxies. Specifically, we consider the lines \nv\ (multiplet), \oip\ (triplet), \silivoiv\ (hereafter simply \silivoivt), \nivd\ (hereafter simply \nivt), \civd\ (hereafter simply \civt), \heii, \oiiid\ (hereafter simply \oiiit), \niii\  (multiplet), \silii\ (multiplet), \siliiid\ (hereafter simply \siliiit), \ciiid\ (hereafter simply \ciiit), \oiiinuv, \neiv, \neiii, \nev\ and \oiid\ (hereafter simply \oiit). We focus on luminosity ratios between emission lines relatively close in wavelength to minimize sensitivity of diagnostics to attenuation by dust \citep{veilleux02}, but we also consider particularly promising diagnostics involving lines separated by several hundred angstr\"om. We exclude the H-Ly$\alpha$ line from our analysis, because of complications linked to the absorption of this resonant line by dust and, at high redshift, circumgalactic neutral hydrogen \citep[e.g.,][]{charlot93,hainline11,schenker14}.

\subsection{Ultraviolet observational samples}\label{sec:UVdata}

To compare the predictions of our models with observations of ultraviolet emission lines in active and inactive galaxies, we appeal to two main samples: (i) a sample of 22 type-2 AGN assembled by \citet[][and references therein]{dors14}, consisting of 12 Seyfert-2 galaxies in the local Universe, with spectra from the \textit{International Ultraviolet Explorer}, and 10 X-ray selected type-2 quasars at redshift $1.5\lesssim z\lesssim4.0$, with VLT/FORS spectra; and (ii) a sample of 5 gravitationally lensed, low-mass star-forming galaxies at redshifts $1.5\lesssim z\lesssim3.0$, with Keck/LRIS and VLT/FORS2 spectra from \cite{stark14}. Line-flux measurements for all these objects are available for \nv, \civt, \heii\ and \ciiit. For star-forming galaxies, measurements are also available for \niii\ and \siliiit. We note that line-flux measurements of AGN narrow-line regions can potentially be contaminated by star formation in the host galaxy \citep[e.g.][]{bonzini13,antonucci15}.

It is important to point out that the above line measurements were not corrected for attenuation by dust. This is not important for the sample of low-mass star-forming galaxies, which have been shown to be extremely dust-poor  \cite[table~7 of][]{stark14}. For the AGN sample, we prefer not to correct the observed flux by adopting an arbitrary attenuation curve, as the dispersion between different standard curves is large at ultraviolet wavelengths \citep[e.g., figure~9 of][]{charlot00}. Instead, when comparing AGN models with these data, we compute the attenuation that would be inferred using the \cite{calzetti00} curve for a $V$-band attenuation of one magnitude ($A_{\rm V}=1$) and mention when this would have a non-negligible effect on our analysis.

\subsection{Diagnostics based on the \civt, \heii\ and \ciiit\ emission lines}\label{sec:CHeC}

\begin{figure*}
\begin{center}
\includegraphics[width=17.0cm]{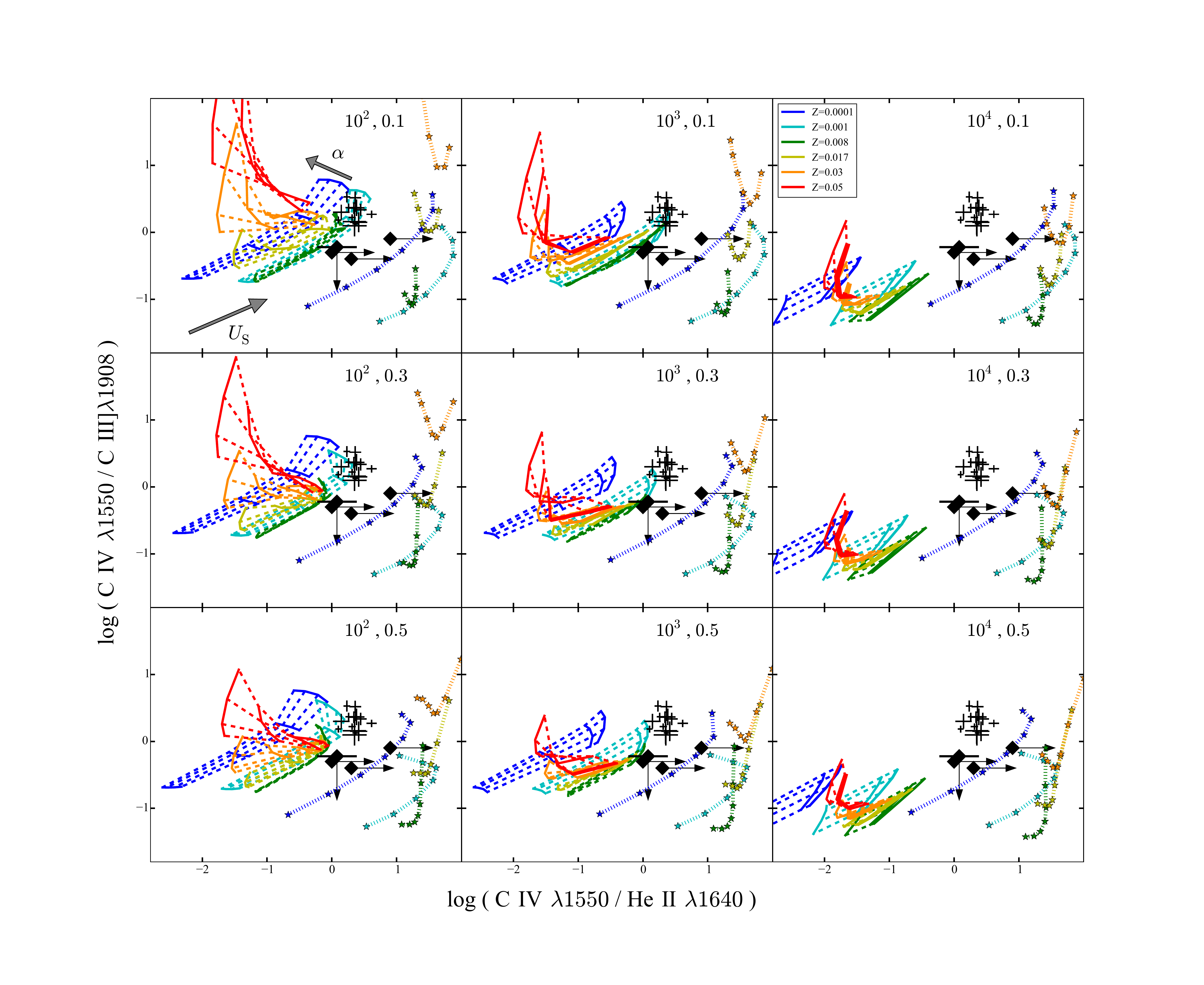} 
\caption{Predictions of the AGN and SF models described in Section~\ref{sec:models} in the \civt/\ciiit\ versus \civt/\heii\ diagnostic diagram, for different assumptions about the gas density ($n_{\rm H}=10^2$, $10^3$ and $10^4$~cm$^{-3}$ from left to right) and dust-to-metal mass ratio ($\xi_{\rm d}=0.1$, 0.3 and 0.5 from top to bottom). In each panel, we show AGN models corresponding to wide ranges in power-law index, $-2.0\leq\alpha\leq-1.2$ (connected by solid lines), and ionization parameter, $-4.0\leq\log U_{\rm S}\leq-1.0$ (connected by dashed lines; $\alpha$ and $U_{\rm S}$ increasing as indicated in the upper-left panel), and SF models (stars connect by dotted lines) corresponding to the same range in $U_{\rm S}$, for different metallicities $Z$  (colour-coded as indicated in the top-right panel). Also shown in each panel are the observations of AGN (crosses with error bars) and star-forming galaxies (large diamonds with upper and lower limits) described in Section~\ref{sec:UVdata}.}
\label{fig:UV_Villar}
\end{center}
\end{figure*}

\begin{figure*}
\begin{center}
\includegraphics[width=17.0cm]{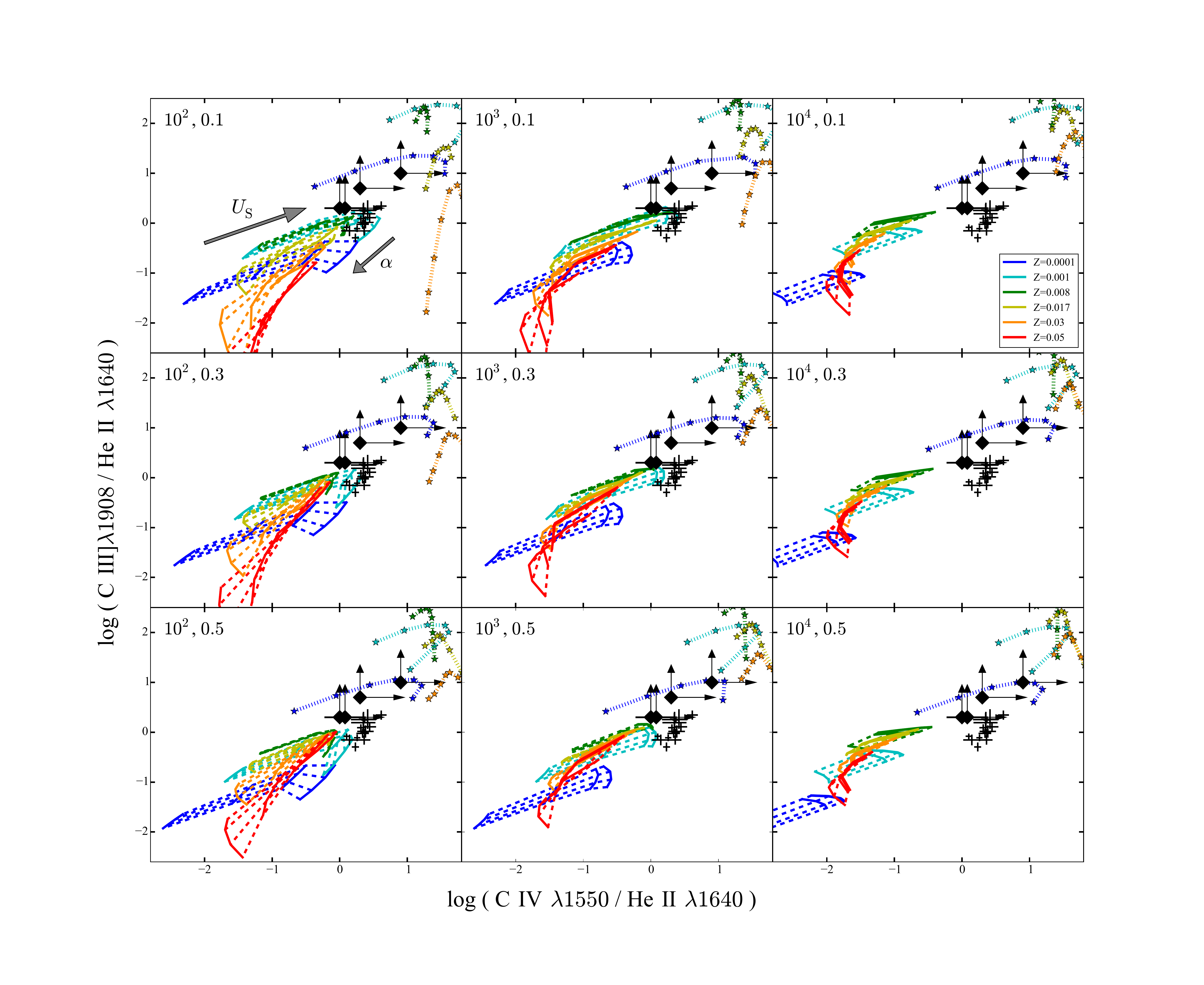} 
\caption{Same as Fig. \ref{fig:UV_Villar}, but for the \ciiit/\heii\ versus \civt/\heii\ diagnostic diagram.}
\label{fig:UV_Villar2}
\end{center}
\end{figure*}

Together with the \heii\ Balmer recombination line, the \civt\ and \ciiit\ collisionally excited line doublets are among the most commonly detected ultraviolet emission lines in galaxy spectra. Diagnostics involving the relative intensities of these 3 lines, originally investigated by \cite{villarmartin97}, are commonly invoked to distinguish between photoionization by AGN and shocks in galaxies. This is because, while AGN and shocks produce similar optical-line ratios, the \civt/\heii\ and \ciiit/\heii\  ultraviolet-line ratios produced by shocks are predicted to be much larger than those produced by AGN \citep[][and references therein]{allen98,groves04b}. These ultraviolet-line ratios have also been used to investigate the metallicity of narrow-line regions of AGN \citep[e.g.,][]{nagao06}. Indeed, combining either \civt/\heii\ or \ciiit/\heii, which are sensitive to metallicity, with \ciiit/\civt, which depends primarily on ionization parameter,  provides good diagnostics of the metallicity in these environments \citep[e.g.][]{groves04b}. In the following paragraphs, we use the AGN and SF photoionization models presented in Section~\ref{sec:models} to examine whether \civt/\heii, \ciiit/\heii\ and \ciiit/\civt\ can help discriminate between active and inactive galaxies.

Figs~\ref{fig:UV_Villar} and \ref{fig:UV_Villar2} show a collection of AGN and SF models in, respectively, the \ciiit/\civt\ versus \civt/\heii\ and \ciiit/\heii\  versus  \civt/\heii\ diagrams. Each panel in these figures corresponds to different assumptions about the gas density, $n_{\rm H}$ (increasing from left to right), and dust-to-metal ratio, $\xi_{\rm d}$ (increasing from top to bottom), as indicated. In each panel, we show AGN models corresponding to wide ranges in power-law index, $\alpha$, and ionization parameter, $U_{\rm S}$, and SF models (stars connect by dotted lines) corresponding to the same range in $U_{\rm S}$, for different metallicities $Z$  (colour-coded as indicated in the top-right panel). Also shown in Figs~\ref{fig:UV_Villar} and \ref{fig:UV_Villar2} are the observations of AGN (crosses with error bars) and star-forming galaxies (large diamonds with upper and lower limits) described in Section~\ref{sec:UVdata}. 

The first impression when looking at Figs~\ref{fig:UV_Villar} and \ref{fig:UV_Villar2} is that models of active and inactive galaxies populate different areas of the \civt/\ciiit\ versus \civt/\heii\ and \civt/\heii\ versus \ciiit/\heii\ diagrams. Furthermore, observations of active galaxies tend to overlap with AGN models (at least for some combinations of parameter), while the upper limits derived from observations of star-forming galaxies point towards SF models. Interestingly, the ultraviolet observations of active galaxies in Figs~\ref{fig:UV_Villar} and \ref{fig:UV_Villar2} tend to favor AGN models with high ionization parameter, $\log U_{\rm S}\sim-1$, while the optical observations of SDSS galaxies in Figs~\ref{fig:BPT1}--\ref{fig:BPT23} were compatible with a much larger range in $U_{\rm S}$  \cite[a similar result was already noted by][]{groves04b}. This finding should be modulated by the fact that the SDSS AGN sample is much larger than that with available ultraviolet data in Figs~\ref{fig:UV_Villar} and \ref{fig:UV_Villar2}. Moreover, the high $U_{\rm S}$ favored by AGN data in these figures could also result, at least in part, from the contamination of line-flux measurements by star formation (Section~\ref{sec:UVdata}). This is compatible with the position of the SF models relative to the AGN models in Figs~\ref{fig:UV_Villar} and \ref{fig:UV_Villar2}.

It is useful to examine how the ultraviolet emission-line properties of AGN models vary as a function of the different adjustable parameters, such as gas density, metallicity, spectral power-law index, ionization parameter and dust-to-metal mass ratio (we refer to G15 for a detailed description of the ultraviolet properties of SF models). According to Figs~\ref{fig:UV_Villar} and \ref{fig:UV_Villar2}, and modulo the above remark about potential contamination by circumnuclear star formation, the models in best agreement with AGN data are those with $n_{\rm H}\lesssim10^3$~cm$^{-3}$ \citep[see also][]{nagao06}. At fixed other parameters, increasing $n_{\rm H}$ means lowering the volume-filling factor, $\epsilon$ (equation \ref{eq:us}).  Since $\epsilon$ scales as $1/\sqrt{n_{\rm H}}$, the dust optical depth $\tau_{\rm d}\propto \xi_{\rm d} Z n_{\rm H} \epsilon$ \citep{brinchmann13} rises as $\sqrt{n_{\rm H}}$. The implied extra absorption of energetic photons causes the electronic temperature, and hence, \civt/\ciiit\ and \civt/\heii, to drop (Fig.~\ref{fig:UV_Villar}). This is also globally the case for \ciiit/\heii\ in Fig.~\ref{fig:UV_Villar2}, except in models with very low ionization parameter ($\log U_{\rm S}\lesssim-3$) and high metallicity ($Z\gtrsim0.017$). We note that, as metallicity increases, the luminosity ratios of both \civt\ and \ciiit\ to \heii\ also increase in Figs~\ref{fig:UV_Villar} and \ref{fig:UV_Villar2} but then stagnate beyond $Z\sim0.008$. As for the optical metal-line transitions discussed in Section~\ref{sec:BPT} (Fig.~\ref{fig:BPT1}), this is caused by a regulation between the rise in the abundance of coolants and the associated drop in electronic temperature, when metallicity increases.

The power-law index of the ionizing spectrum, $\alpha$, controls the relative proportions of photons capable of ionizing different elements. As shown by Fig.~\ref{fig:ion_spec}, the ionization potential rises from C\,{\sc iii}, to He\,{\sc ii}, to C\,{\sc iv}. Thus, at fixed other parameters, we expect \ciiit/\heii\ to drop and both \civt/\ciiit\ and \civt/\heii\ to rise, when $\alpha$ increases from $-2.0$ (steepest AGN spectrum in Fig.~\ref{fig:ion_spec}) to $-1.2$ (flattest spectrum). This is globally what Figs~\ref{fig:UV_Villar} and \ref{fig:UV_Villar2} show, except for some combinations of parameters. For example, a rise in $\alpha$ appears to trigger a drop in \civt/\ciiit\ at high metallicity ($Z\gtrsim0.017$; Fig.~\ref{fig:UV_Villar}) and a drop in \civt/\heii\ at low gas density ($n_{\rm H}=10^2$~cm$^{-3}$; Figs~\ref{fig:UV_Villar} and \ref{fig:UV_Villar2}). Also, increasing the ionization parameter $U_{\rm S}$, which corresponds here to increasing the filling factor $\epsilon$ at fixed $\alpha$ and gas density $n_{\rm H}$ (equation \ref{eq:us}), causes the ionized nebula to be more compact and concentrated close to the ionizing source (equation \ref{eq:rs}). This makes \civt/\ciiit, \civt/\heii\ and \ciiit/\heii\  increase with the ionization parameter, $U_{\rm S}$, except for the drop in \civt/\ciiit\ in the most metal-rich models (Fig.~\ref{fig:UV_Villar}). Fig.~\ref{fig:UV_Villar2} further shows that \civt/\heii\ and \ciiit/\heii\ never exceed $\sim0.6\,$dex in AGN models, in agreement with results of \cite{groves04b}.

Changes in the dust-to-metal mass ratio have a complex effect on \civt/\ciiit, \civt/\heii\ and \ciiit/\heii\ in the AGN models in Figs~\ref{fig:UV_Villar} and \ref{fig:UV_Villar2}, which results from the combination of several factors. An increase in $\xi_{\rm d}$ causes the electronic temperature to rise because of the depletion of coolants from the gas phase, which should increase \civt/\heii\ and \ciiit/\heii\ (for reference, about 60 per cent of carbon is depleted onto dust grains for $\xi_{\rm d}=0.5$). At the same time, fewer C ions can be collisionally excited in the gas phase, which should lower \civt/\heii\ and \ciiit/\heii. Moreover, increasing $\xi_{\rm d}$ raises the dust optical depth in the ionized gas, which scales as $\tau_{\rm d}\propto \xi_{\rm d} Z n_{\rm H} \epsilon$ \citep{brinchmann13}. The effect is accentuated at high ionization parameter, since the filling factor scales as $\epsilon \propto U_{\rm S}^{3/2}$ at fixed other parameters (equation \ref{eq:us}). The predictions in Figs~\ref{fig:UV_Villar} and \ref{fig:UV_Villar2} therefore require a careful self-consistent treatment of metal abundances and depletion, such as that described in Section~\ref{sec:models} above (see also G15).

The position of AGN models (and observations) versus SF ones in Figs~\ref{fig:UV_Villar} and \ref{fig:UV_Villar2} provides valuable information about the potential of \civt/\ciiit, \civt/\heii\ and \ciiit/\heii\ to discriminate between active and inactive galaxies. For example, the similar ranges in \civt/\ciiit\ spanned by AGN and SF models in Fig.~\ref{fig:UV_Villar} indicate that knowledge of this line ratio alone is not enough distinguish between the 2 types of photoionization. In contrast, as Fig.~\ref{fig:UV_Villar2} shows, \civt/\heii\ and \ciiit/\heii\ can help separate, each on their own, between nuclear activity and star formation, except in a few extreme cases: in the most metal-rich SF models with $n_{\rm H}=10^{2}$~cm$^{-3}$ and $\xi_{\rm d}\lesssim0.3$, \ciiit/\heii\ can resemble that of AGN models, as can \civt/\heii\ in the most metal-poor SF models with $\log U_{\rm S}\lesssim-3$. In this context, combined information about \civt/\heii\ and \ciiit/\heii\ provides more unequivocal constraints on the nature of the ionizing source, while additional information on \civt/\ciiit\ will provide further clues on the metal and dust abundance, ionization parameter and spectral power-law index (Fig.~\ref{fig:UV_Villar}).

In the next subsections, we investigate emission-line ratio based on other ion species. For this purpose, it is useful to define a reference set of model parameters. As in Section~\ref{sec:others} above, we adopt for simplicity a single choice of gas density, which differs for the AGN models, $n_{\rm H}=10^3$~cm$^{-3}$, and the SF models, $n_{\rm H}=10^2$~cm$^{-3}$. Also, except when otherwise indicated, we adopt a single dust-to-metal mass ratio, $\xi_{\rm d}=0.3$, in the middle of the explored range. Finally, in all line-ratio diagrams in the remainder of this section, we adopt \ciiit/\heii\ as the ordinate. It is because this ratio has been shown above to be a good discriminant between photoionization by an AGN and star formation, and hence, it will allow a straightforward identification of other line ratios sensitive to the nature of the ionizing source. We favour \ciiit/\heii\ over \civt/\heii\ in this exercise, because observations of \civt\ can be contaminated by the P-Cygni profile of photospheric stellar emission in young galaxies, while strong \ciiit\ emission is now commonly detected in high-redshift galaxies \citep[e.g.,][]{stark15a,zitrin15}.

\subsection{\nv-based diagnostics}

\begin{figure*}
\begin{center}
\includegraphics[width=17.0cm]{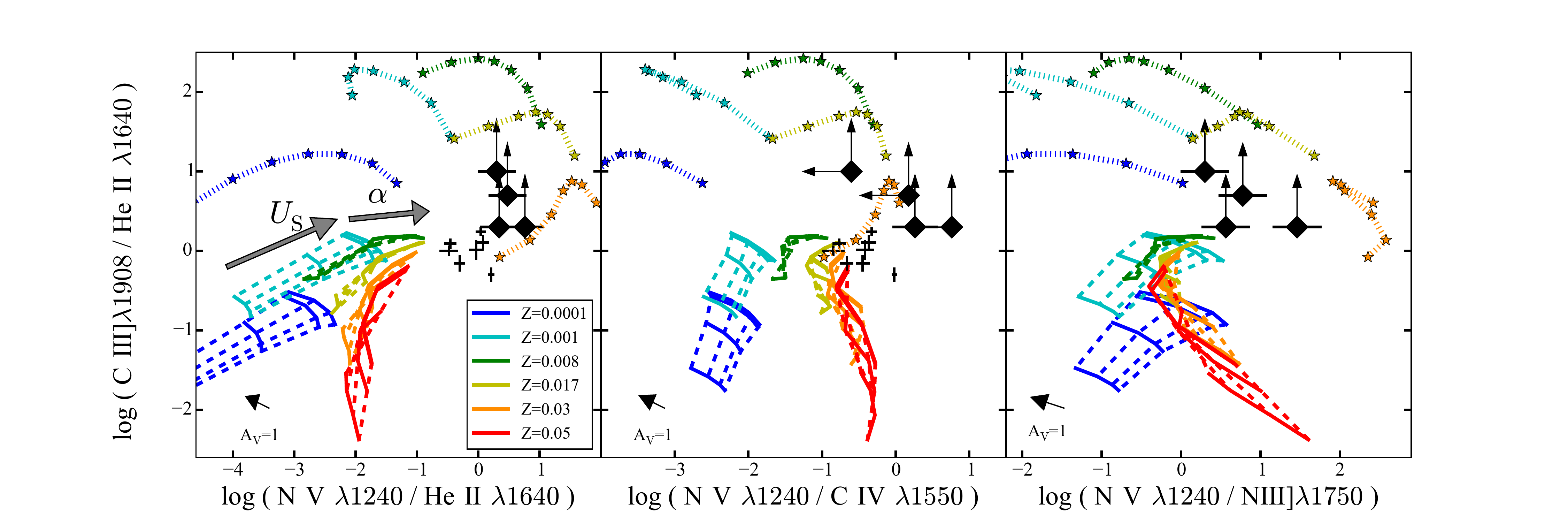} 
\caption{Predictions of the AGN and SF models described in Section~\ref{sec:models} in the \ciiit/\heii\ versus \nv/\heii, \nv/\civt\ and \nv/\niii\ diagnostic diagrams (from left to right), for a dust-to-metal mass ratio $\xi_{\rm d}=0.3$ (our reference value in Section~\ref{sec:CHeC}). AGN and SF models have gas densities $n_{\rm H}=10^3$ and $10^2$~cm$^{-3}$, respectively. In each panel, we show AGN models corresponding to wide ranges in power-law index, $-2.0\leq\alpha\leq-1.2$ (connected by solid lines), and ionization parameter, $-4.0\leq\log U_{\rm S}\leq-1.0$ (connected by dashed lines; $\alpha$ and $U_{\rm S}$ increasing as indicated in the left panel), and SF models (stars connect by dotted lines) corresponding to the same range in $U_{\rm S}$, for different metallicities $Z$  (colour-coded as indicated in the left panel). Also shown in each panel are the observations of AGN (crosses with error bars) and star-forming galaxies (large diamonds) described in Section~\ref{sec:UVdata}, when available. In each panel, black arrows indicate the effect of attenuation by dust for $A_{\rm V}=1\,$mag and a \protect\citet{calzetti00} attenuation curve (AGN data are not corrected for attenuation).}
\label{fig:UV_NV}
\end{center}
\end{figure*}

\begin{figure*}
\begin{center}
\includegraphics[width=17.0cm]{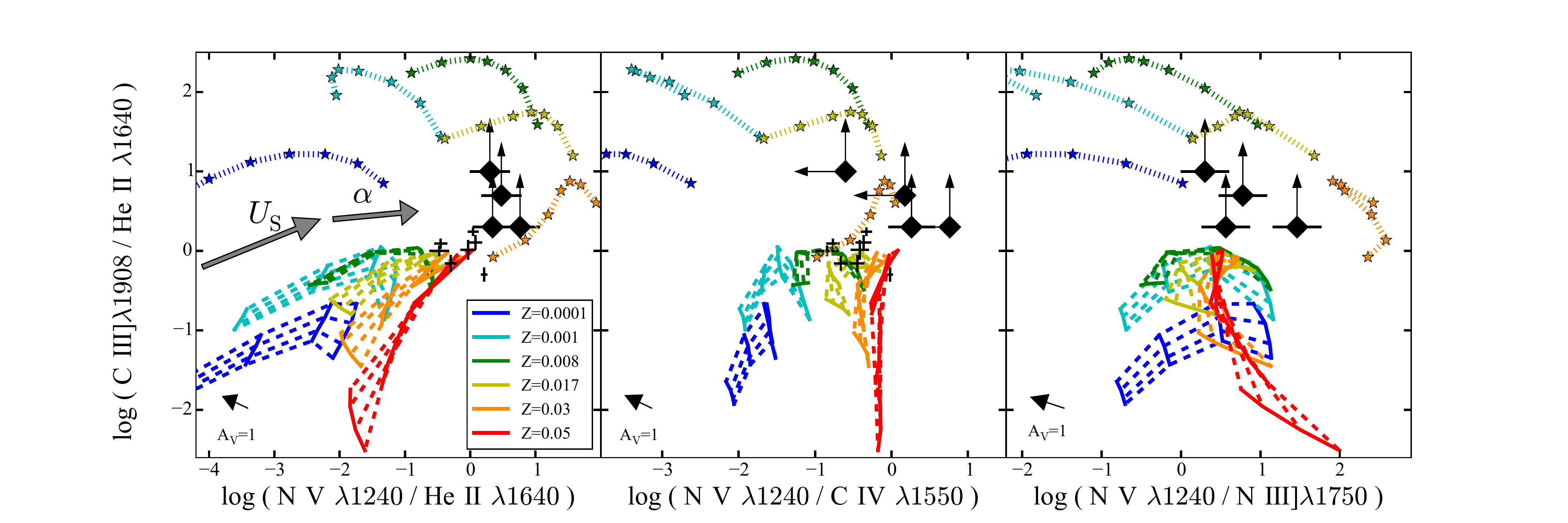} 
\caption{Same as Fig.~\ref{fig:UV_NV}, but for a dust-to-metal mass ratio $\xi_{\rm d}=0.5$ and an AGN model gas density $n_{\rm H}=10^2$~cm$^{-3}$.}
\label{fig:UV_NV_ter}
\end{center}
\end{figure*}

Ultraviolet emission-line ratios involving the \nv\ multiplet, such as \nv/\civt\ and \nv/\heii, have been exploited to interpret observations of active galaxies using AGN photoionization models, in particular as metallicity diagnostics \citep[][and references therein]{groves04b,nagao06}. Here, we investigate the extent to which \nv-based emission-line ratios can be used as diagnostics of the nature of the photoionization source in a galaxy.

In Fig.~\ref{fig:UV_NV}, we show \ciiit/\heii\ as a function of \nv/\heii, \nv/\civt\ and \nv/\niii\ (from left to right) for a collection of AGN and SF models with reference values of the gas density $n_{\rm H}$ and dust-to-metal mass ratio $\xi_{\rm d}$ (Section~\ref{sec:CHeC}). In each panel, we show AGN and SF models corresponding to same ranges in power-law index, $\alpha$, ionization parameter, $U_{\rm S}$, and metallicity, $Z$, as in Fig.~\ref{fig:UV_Villar}. Also shown in Fig.~\ref{fig:UV_NV} are the observations of AGN (crosses) and star-forming galaxies (large diamonds) described in Section~\ref{sec:UVdata}, when available. We draw the dust attenuation vector in these figures, which, unlike in Figs~\ref{fig:UV_Villar} and \ref{fig:UV_Villar2}, is not negligible because of the wide wavelength range spanned by \nv-based line-ratio diagnostics (we recall that AGN data are not corrected for attenuation; see Section~\ref{sec:UVdata}).

Fig.~\ref{fig:UV_NV} shows that AGN and SF models populate distinct areas of the \ciiit/\heii\ versus \nv/\heii, \nv/\civt\ and \nv/\niii\  diagrams, exhibiting the same trends as the observations of active and inactive galaxies. We note that AGN models around solar metallicity have traditionally been found to predict lower \nv\ line luminosity than observed in broad- and narrow-line regions of AGN \citep[e.g.,][]{osmer76,kraemer98}. In fact, some authors appeal to very high metallicities (up to $Z\sim10Z_{\odot}$) to reproduce observations \citep[][]{netzer13}. We show in Fig.~\ref{fig:UV_NV_ter} that by choosing slightly different values of the dust-to-metal ratio, $\xi_{\rm d}=0.5$, and gas density in narrow-line regions, $n_{\rm H}=10^2$~cm$^{-3}$, our AGN models are in better agreement with available observations of active galaxies. In this context, we should also keep in mind the possible contamination of these nuclear line-flux measurements by star formation in Figs~\ref{fig:UV_NV} and \ref{fig:UV_NV_ter} (Section~\ref{sec:UVdata}).

The behaviour of models with different $n_{\rm H}$, $\alpha$, $U_{\rm S}$ and $\xi_{\rm d}$ in Figs~\ref{fig:UV_NV} and \ref{fig:UV_NV_ter} can be largely understood from that described for the diagrams defined by the \civt, \heii\ and \ciiit\ emission lines in Figs~\ref{fig:UV_Villar} and \ref{fig:UV_Villar2}. For example, increasing the gas density raises the dust optical depth, which lowers the electronic temperature, and hence, \nv/\heii, \nv/\civt\ and \nv/\niii. Also, a steepening of the ionizing spectrum (from $\alpha=-1.2$ to $-2.0$) reduces the amount of highly energetic photons, which should make \nv/\heii, \nv/\civt\ and \nv/\niii\  drop according to the ionization potentials of these ions (Fig.~\ref{fig:ion_spec}). At high $Z$ and high $\xi_{\rm d}$, the effect in Fig.~\ref{fig:UV_NV_ter} is modulated by the increased amount of dust and the different depletion levels of carbon and nitrogen. As noted previously (Section~\ref{sec:CHeC}), raising the ionization parameter $U_{\rm S}$ at fixed $\alpha$ and $n_{\rm H}$ makes the ionized nebula more concentrated close to the ionizing source. This causes \nv/\heii, \nv/\civt\ and \nv/\niii\ to globally increase together with \ciiit/\heii\ in Figs~\ref{fig:UV_NV} and \ref{fig:UV_NV_ter}, although the changes are more complex at the highest metallicities for \nv/\civt\ and \nv/\niii. An increase in $\xi_{\rm d}$ causes a depletion of coolants from the gas phase, a rise in the electronic temperature, and hence, since nitrogen is a non-refractory element, a rise in \nv/\heii.

Figs~\ref{fig:UV_NV} and \ref{fig:UV_NV_ter} indicate that \nv/\heii\, \nv/\civt\ and \nv/\niii\  cannot be used individually to discriminate between photoionization by an AGN and star formation in a galaxy, even though positive values of $\log( \nv/\heii)$ appear to be achievable only with metal-rich stellar ionizing spectra. This is consistent with the fact that such spectra differ most markedly from AGN spectra (Fig.~\ref{fig:ion_spec}). Combined with information about \ciiit/\heii, \nv-based line ratios can provide interesting clues about the nature of the ionizing source, and even about metallicity and the ionization parameter in galaxies powered by star formation. Yet, \nv-based diagnostics are not optimal because of their strong dependence on the details of secondary nitrogen enhancement. Also, as indicated by the large dynamic range spanned by the abscissae in Figs~\ref{fig:UV_NV} and \ref{fig:UV_NV_ter}, \nv\ emission is very weak and hence difficult to detect at low metallicity \citep{nagao06}.

\subsection{\heii-based diagnostics}\label{sec:he}

\begin{figure*}
\begin{center}
\includegraphics[width=17.0cm]{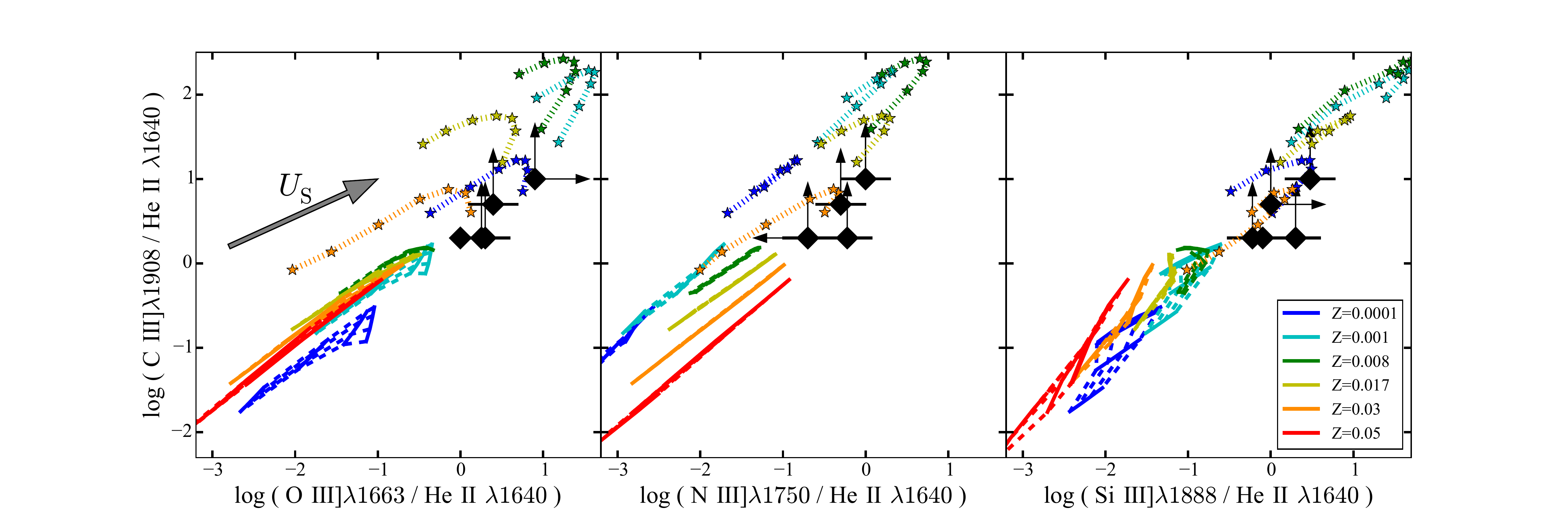} 
\caption{Predictions of the AGN and SF models described in Section~\ref{sec:models} in the \ciiit/\heii\ versus \oiiit/\heii, \niii/\heii\ and \siliiit/\heii\ diagnostic diagrams (from left to right), for a dust-to-metal mass ratio $\xi_{\rm d}=0.3$ and gas densities $n_{\rm H}=10^3$ and $10^2$~cm$^{-3}$ for AGN and SF models, respectively (our reference values in Section~\ref{sec:CHeC}). In each panel, the models and observations (when available) are the same as in Fig.~\ref{fig:UV_NV}.}
\label{fig:UV_HeII}
\end{center}
\end{figure*}

The \heii\ recombination line depends less strongly on metallicity and ionization parameter than collisionally excited metal lines. For this reason, \heii\ can be used as a standard reference line at ultraviolet wavelengths, similarly to H$\beta$ at optical wavelengths. In this section, we propose 3 \heii-based emission-line ratio diagnostics to distinguish between photoionization by an active nucleus and star formation. Specifically, we show in Fig.~\ref{fig:UV_HeII} the  \ciiit/\heii\ versus \oiiit/\heii, \niii/\heii\ and \siliiit/\heii\ diagrams (from left to right) for the same collection of AGN and SF models as in Fig.~\ref{fig:UV_NV}.

AGN and SF models occupy distinct areas of the 3 diagrams in Fig.~\ref{fig:UV_HeII}, the observations of low-mass star-forming galaxies being consistent with the predictions of SF models. Thus, \oiiit/\heii, \niii/\heii\ and \siliiit/\heii\ could be used individually to constrain the dominant source of ionizing photons in a galaxy. We note that, although the most metal-rich SF models ($Z=0.03$) can reach values of \oiiit/\heii, \niii/\heii\ and \siliiit/\heii\  similar to those in some AGN models, this should not weaken the value of these diagnostic line ratios to interpret observations of only moderately enriched galaxies at high redshifts. Fig.~\ref{fig:UV_HeII} further shows that, while AGN models with different metallicities separate well in the \ciiit/\heii\ versus  \niii/\heii\ diagram, this is not the case for the \ciiit/\heii\ versus \oiiisf/\heii\ and \siliiit/\heii\ diagrams. Also, none of the diagrams in this figure can help constrain the power-law index $\alpha$ of the ionizing radiation. It is worth pointing out that, as shown in Appendix~\ref{app:sep}, the predictions in the left panel of Fig.~\ref{fig:UV_HeII} are similar to those that would be obtained using only one component (at either $\lambda=1661$ or 1666 \AA) of the \oiiisf\ doublet.

\subsection{O-based diagnostics in the far and near ultraviolet}

\begin{figure*}
\begin{center}
\includegraphics[width=17.0cm]{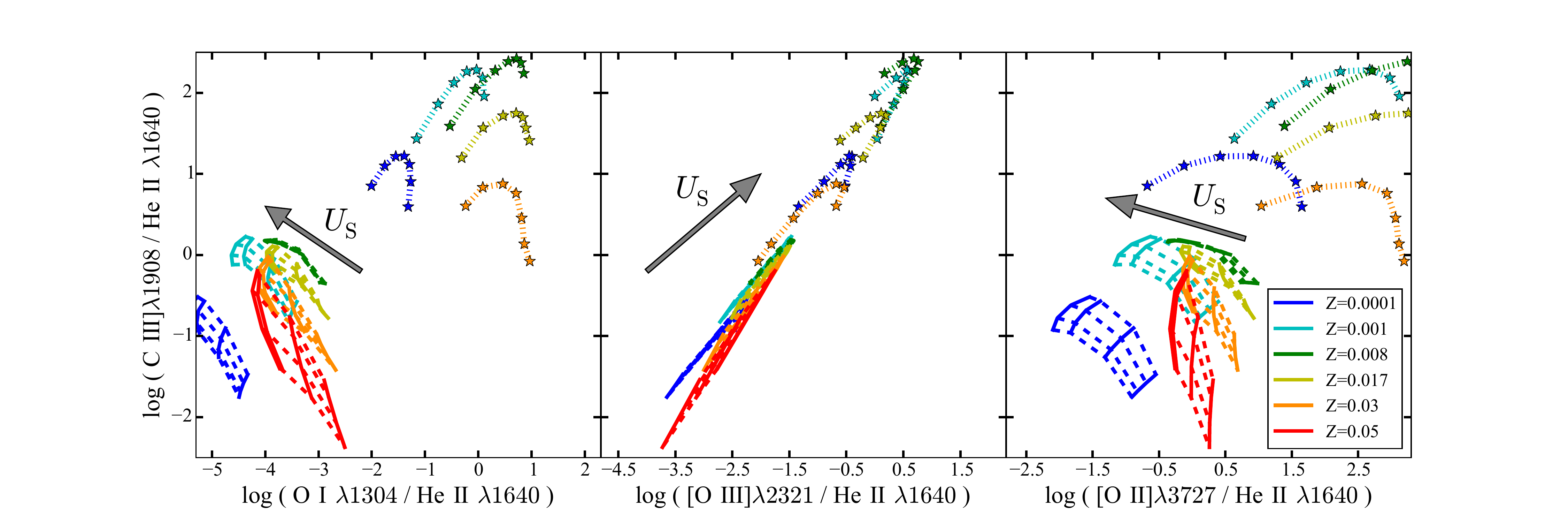} 
\caption{Predictions of the AGN and SF models described in Section~\ref{sec:models} in the \ciiit/\heii\ versus \oip/\heii, \oiiinuv/\heii\ and  \oiit/\heii\ diagnostic diagrams (from left to right), for a dust-to-metal mass ratio $\xi_{\rm d}=0.3$ and gas densities $n_{\rm H}=10^3$ and $10^2$~cm$^{-3}$ for AGN and SF models, respectively (our reference values in Section~\ref{sec:CHeC}). In each panel, the models are the same as in Fig.~\ref{fig:UV_NV}.}
\label{fig:UV_Oxy}
\end{center}
\end{figure*}

In the previous section, we have introduced \oiiit/\heii\ as an ultraviolet emission-line diagnostic of nuclear activity versus star formation in galaxies. Other oxygen transitions, such as \oip, \oiiinuv\ and \oiit\ can also potentially help discriminate between photoionization by an AGN and star formation when combined with \heii.

Fig.~\ref{fig:UV_Oxy} shows \ciiit/\heii\ as a function of \oip/\heii, \oiiinuv/\heii\ and \oiit/\heii\  (from left to right) for the same collection of AGN and SF models as in Fig.~\ref{fig:UV_NV}, with reference values of the gas density $n_{\rm H}$ and dust-to-metal mass ratio (Section~\ref{sec:CHeC}), in wide ranges of power-law index $\alpha$ and ionization parameter $U_{\rm S}$. AGN and SF models occupy distinct regions of these diagrams,  in agreement with available observations of star-forming galaxies. This makes different combinations of the line ratios in Fig.~\ref{fig:UV_Oxy} good diagnostics of nuclear activity versus star formation. We note the similarity of \oiiisf/\heii\ in AGN models and the most metal-rich SF models (middle panel) and of \oiit/\heii\ in AGN models and the most metal-poor SF models (right panel).

It is also worth noting that, given the wide predicted dynamic range in \oip/\heii\ and \oiiinuv/\heii\ in Fig.~\ref{fig:UV_Oxy}, simple lower and upper limit on these emission-line ratios can already provide valuable clues about the nature of the ionizing source in a galaxy. For example, upper limits such as  $\log(\oip/\heii)\lesssim-2.5$ and $\log(\oiiinuv/\heii)\lesssim-2.5$ could be enough to identify the presence of an AGN. We further note that attenuation by dust should have a negligible effect on the capacity to discriminate between nuclear activity and star formation using the diagrams in Fig.~\ref{fig:UV_Oxy}. For example, adopting the \cite{calzetti00} curve, an attenuation of $A_{\rm V}=1$ would affect O-based line ratios by between 0.1 and 0.4\,dex and \ciiit/\heii\ by only 0.07\,dex. A potentially more serious complication is that, in AGN spectra, the \oip\ emission-line triplet is expected to be blended with the \siip\ doublet \citep[e.g.][]{matsuoka07}, although \citet{rodriguez-ardila02} manage to robustly de-blend these components in high-S/N \textit{HST} spectroscopic observations of a Seyfert 1 galaxy.

\subsection{Ne-based diagnostics in the near ultraviolet}\label{sec:Ne-based}

\begin{figure*}
\begin{center}
\includegraphics[width=17.0cm]{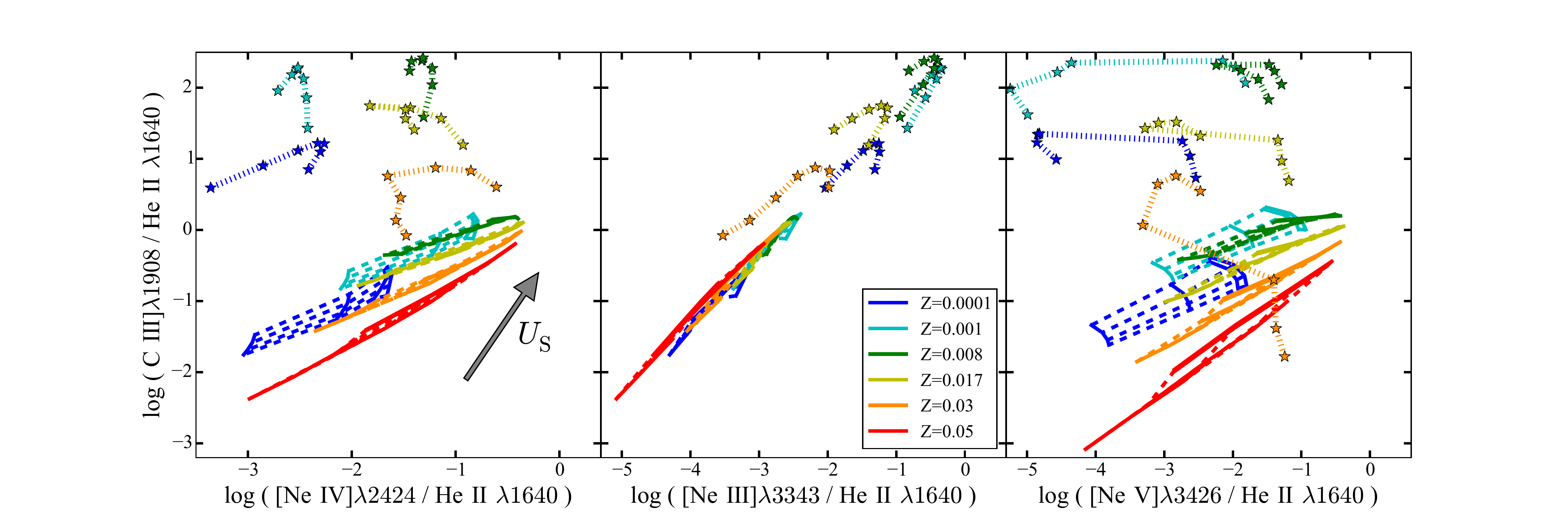} 
\caption{Predictions of the AGN and SF models described in Section~\ref{sec:models} in the \ciiit/\heii\ versus \neiv/\heii, \neiii/\heii\ and \nev/\heii\ diagnostic diagrams (from left to right), for a dust-to-metal mass ratio $\xi_{\rm d}=0.3$ and gas densities $n_{\rm H}=10^3$ and $10^2$~cm$^{-3}$ for AGN and SF models, respectively (our reference values in Section~\ref{sec:CHeC}). In each panel, the models are the same as in Fig.~\ref{fig:UV_NV}.}
\label{fig:UV_Ne}
\end{center}
\end{figure*}

\begin{figure*}
\begin{center}
\includegraphics[width=17.0cm]{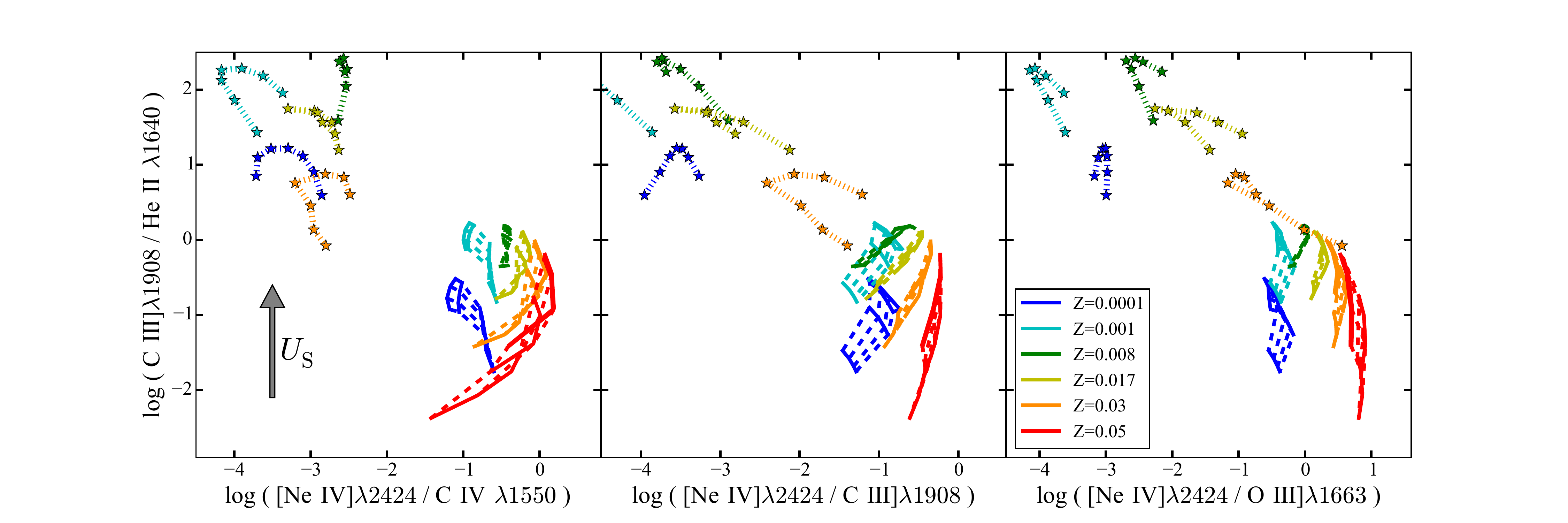}\\
\includegraphics[width=17.0cm]{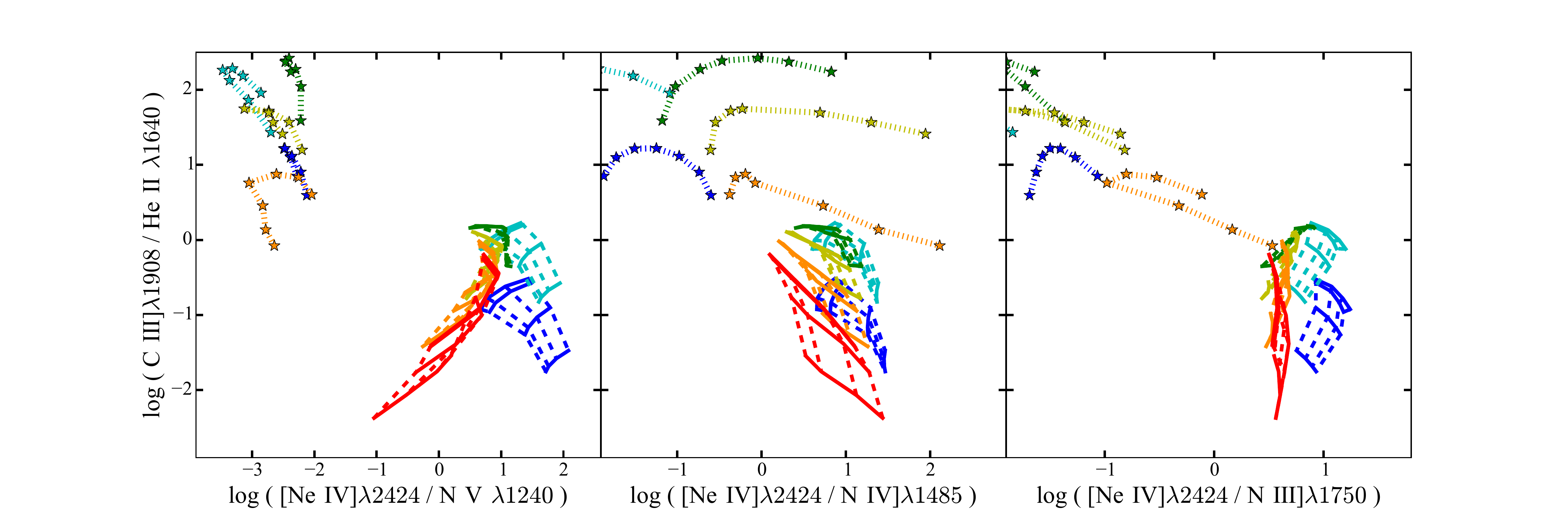} \\
\includegraphics[width=17.0cm]{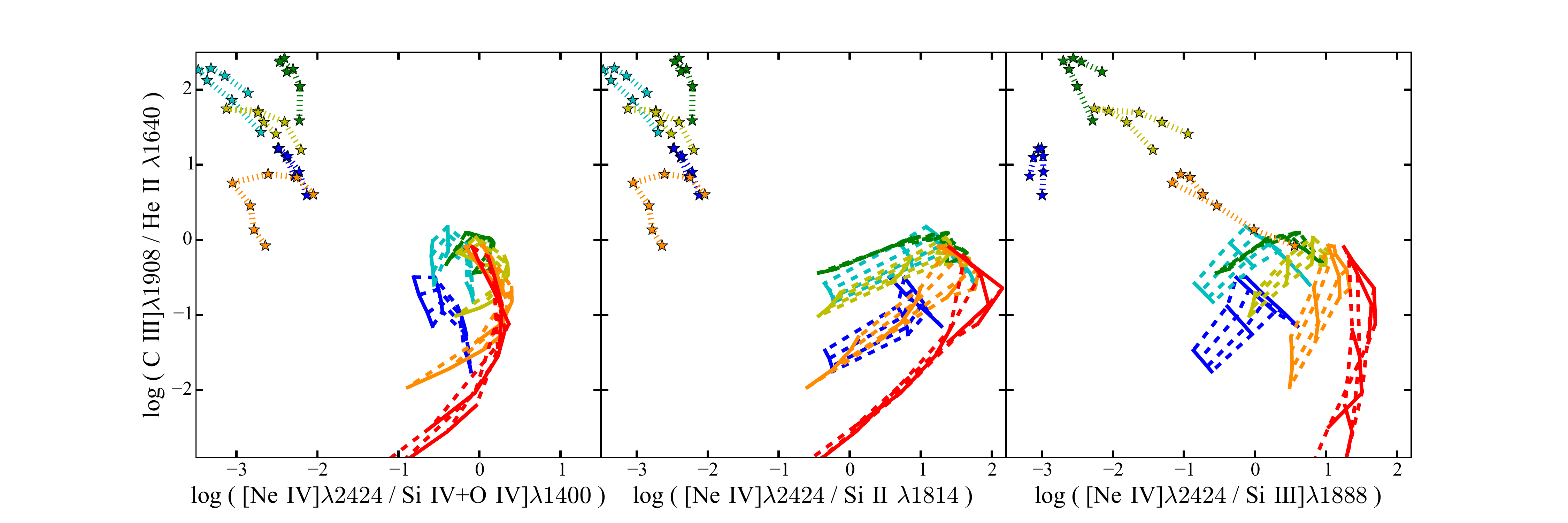} \\
\caption{Predictions of the AGN and SF models described in Section~\ref{sec:models} in several diagnostic diagrams defined by  \ciiit/\heii\ against various \neiv-based line ratios, for a dust-to-metal mass ratio $\xi_{\rm d}=0.3$ and gas densities $n_{\rm H}=10^3$ and $10^2$~cm$^{-3}$ for AGN and SF models, respectively (our reference values in Section~\ref{sec:CHeC}). In each panel, the models are the same as in Fig.~\ref{fig:UV_NV}.}
\label{fig:UV_Ne4}
\end{center}
\end{figure*}

\begin{figure*}
\begin{center}
\includegraphics[width=17.0cm]{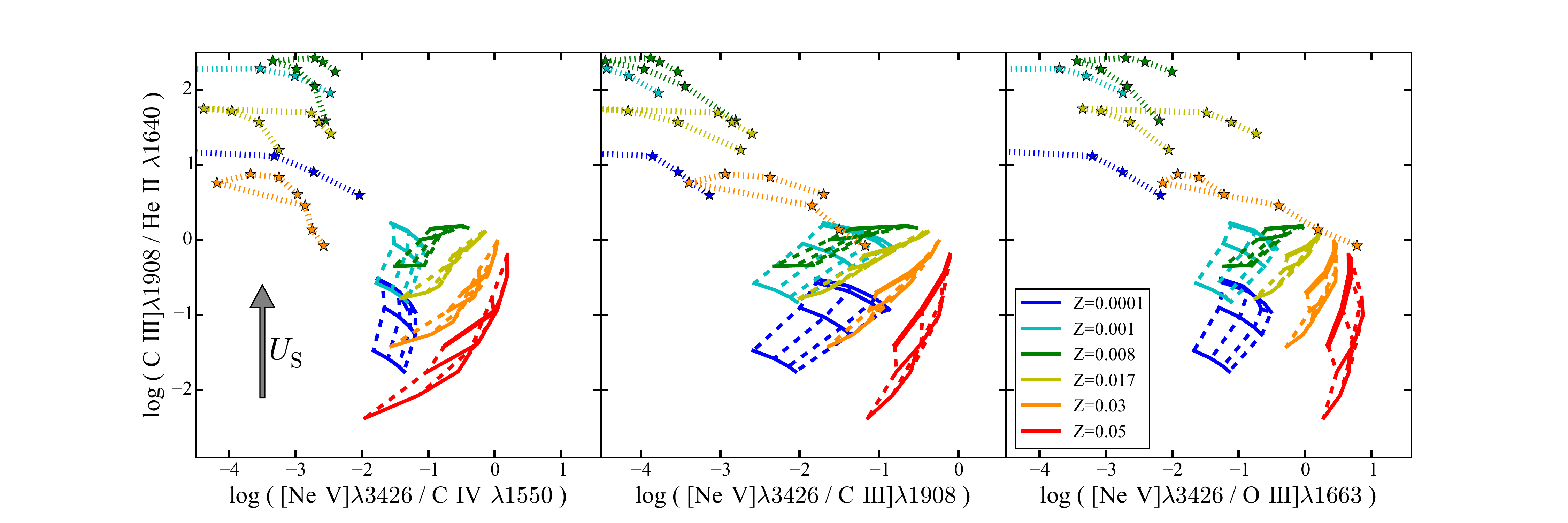} \\
\includegraphics[width=17.0cm]{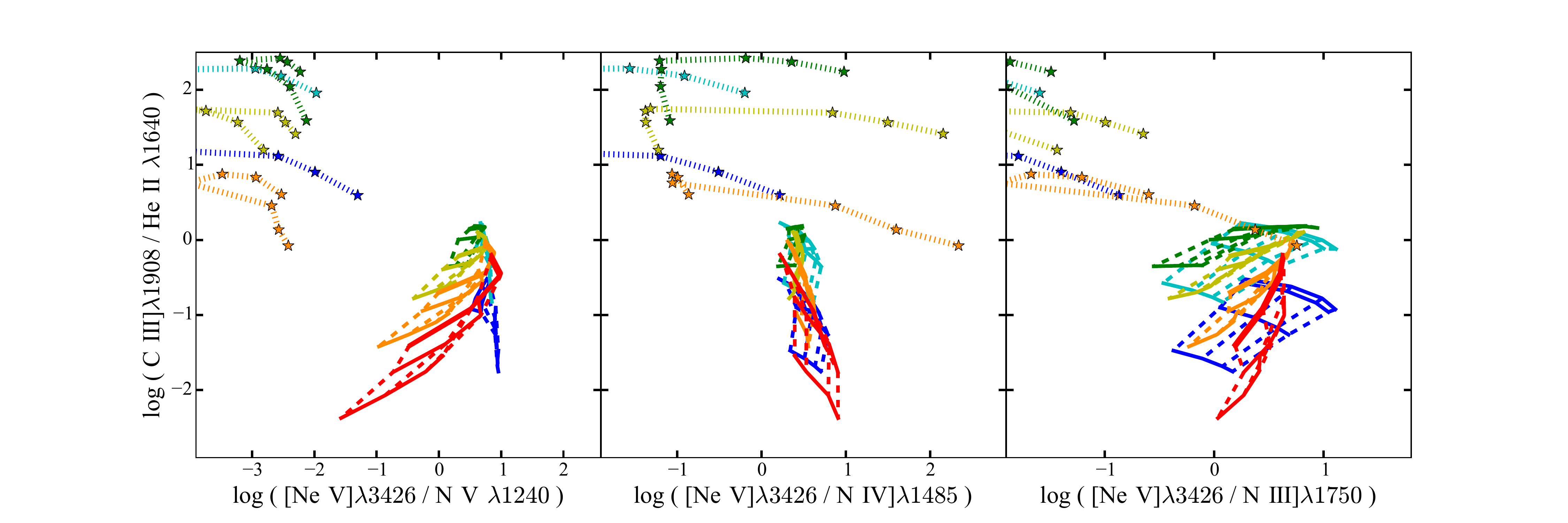} \\
\includegraphics[width=17.0cm]{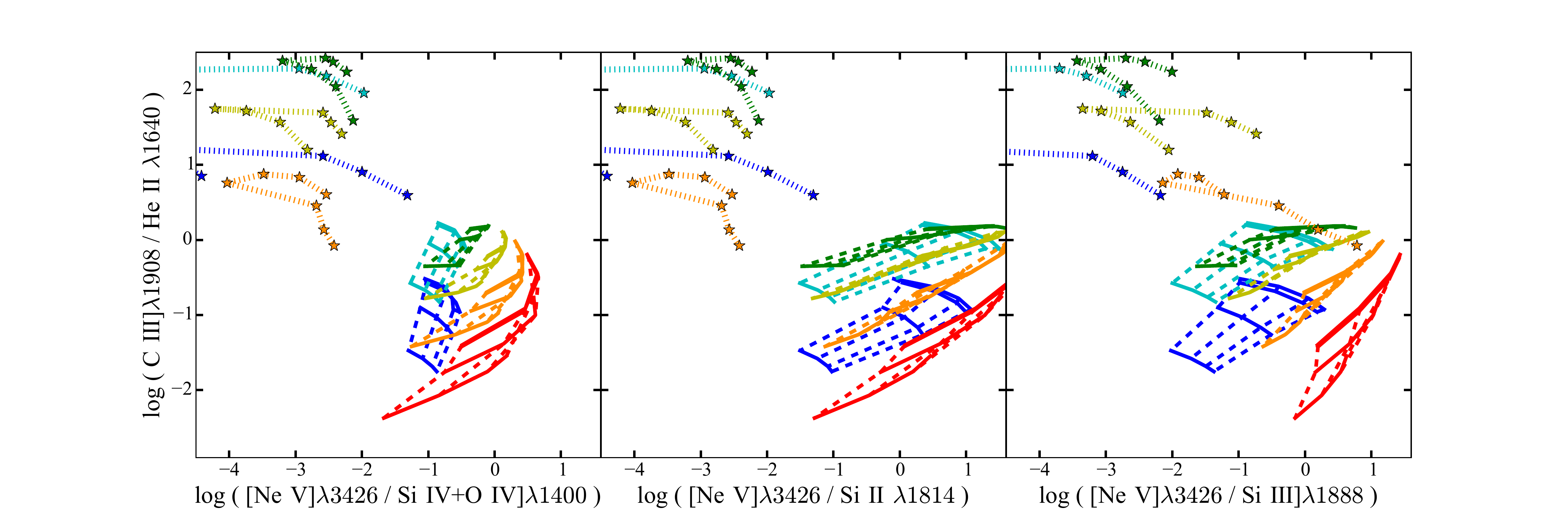}
\caption{Same as Fig.~\ref{fig:UV_Ne4}, but for \nev-based instead of \neiv-based line ratios.}
\label{fig:UV_Ne5}
\end{center}
\end{figure*}

The high ionization potential of Ne{\sc v} ($\sim97 \,{\rm eV}$) requires a hard ionizing spectrum for significant \nev\ emission to be produced, which is unlikely to arise from stellar populations (see Fig.~\ref{fig:ion_spec}). For this reason, a detection of \nev\  has been claimed to be a powerful diagnostic of the presence of an AGN \citep[][and references therein]{mignoli13}. We now explore several Ne-based line-ratio diagnostics involving the \neiii, \neiv\ and \nev\ emission lines (given the faintness of [NeV] emission, we consider here only \nev\ and not the $\sim3$ times weaker [Ne\,{\sc v}]$\lambda3346$ line; \citealt{vandenberk01}). 

Fig.~\ref{fig:UV_Ne} shows \ciiit/\heii\ as a function of \neiv/\heii,  \neiii/\heii\ and \nev/\heii\ (from left to right) for the same collection of AGN and SF models as in Fig.~\ref{fig:UV_NV}. As expected from Fig.~\ref{fig:ion_spec}, AGN models reach the highest \neiv/\heii\ and especially \nev/\heii\ values, but they overlap with SF models for lower values of these line ratios. Interestingly, AGN and SF models separate more markedly in the  \ciiit/\heii\  versus \neiii/\heii\ diagram (middle panel of Fig.~\ref{fig:UV_Ne}). 

To further explore the usefulness of \neiv\ and \nev\ to discriminate between nuclear activity and star formation, we compute luminosity ratios between these emission lines and several lines from other ions. In Figs~\ref{fig:UV_Ne4} and \ref{fig:UV_Ne5}, we show \ciiit/\heii\ as a function of the luminosity ratios defined by \neiv\ and \nev, respectively, and each of the following 9 emission lines: \civt, \ciiit, \oiiit, \nv, \nivt, \niii, \silivoivt, \silii\ and \siliiit. The clear separation between active and inactive galaxies in these diagrams is remarkable. It indicates that combinations of \ciiit/\heii\ with any of the explored Ne-based line ratios can help discriminate between nuclear activity and star formation. Moreover, several Ne-based emission-line ratios investigated in Figs~\ref{fig:UV_Ne4} and \ref{fig:UV_Ne5} are, on their own, good diagnostics of photoionization by an AGN versus star formation. This is the case for \neiv/\civt, \nev/\civt, \neiv/\ciiit, \neiv/\nv, \nev/\nv, \neiv/\niii, \neiv/\silivoivt, \nev/\silivoivt, \neiv/\silii, \nev/\silii\ and \neiv/\siliiit. The other Ne-based emission-line ratios in Figs~\ref{fig:UV_Ne4} and \ref{fig:UV_Ne5} tend to reach, in the most metal-rich SF models, values similar to those of AGN models, requiring \ciiit/\heii\ information for a more secure identification of the ionizing source.

We note that, although the diagnostics diagrams in Figs~\ref{fig:UV_Ne4} and \ref{fig:UV_Ne5} were constructed using dust-free models, the dynamic range in Ne-based line-ratio strength is so large that attenuation by dust should not affect significantly the above results. For example, adopting the \cite{calzetti00} curve, an attenuation of $A_{\rm V}=1$ would affect Ne-based line ratios by between 0.1 and 0.5\,dex (from the smallest to the largest wavelength leverage) and \ciiit/\heii\ by only 0.07\,dex.  More problematic could be the challenge of measuring some of these lines. For example, \neiii\ can potentially be blended with the \nevbis\ line \citep{schirmer13}, while \silii\ has been found to be very weak in high-redshift quasars \citep{negrete14}. 

\subsection{Distinguishing active from inactive galaxies in emission line-ratio diagrams}\label{sec:separation}

\begin{figure}
\begin{center}
\includegraphics[width=8.0cm]{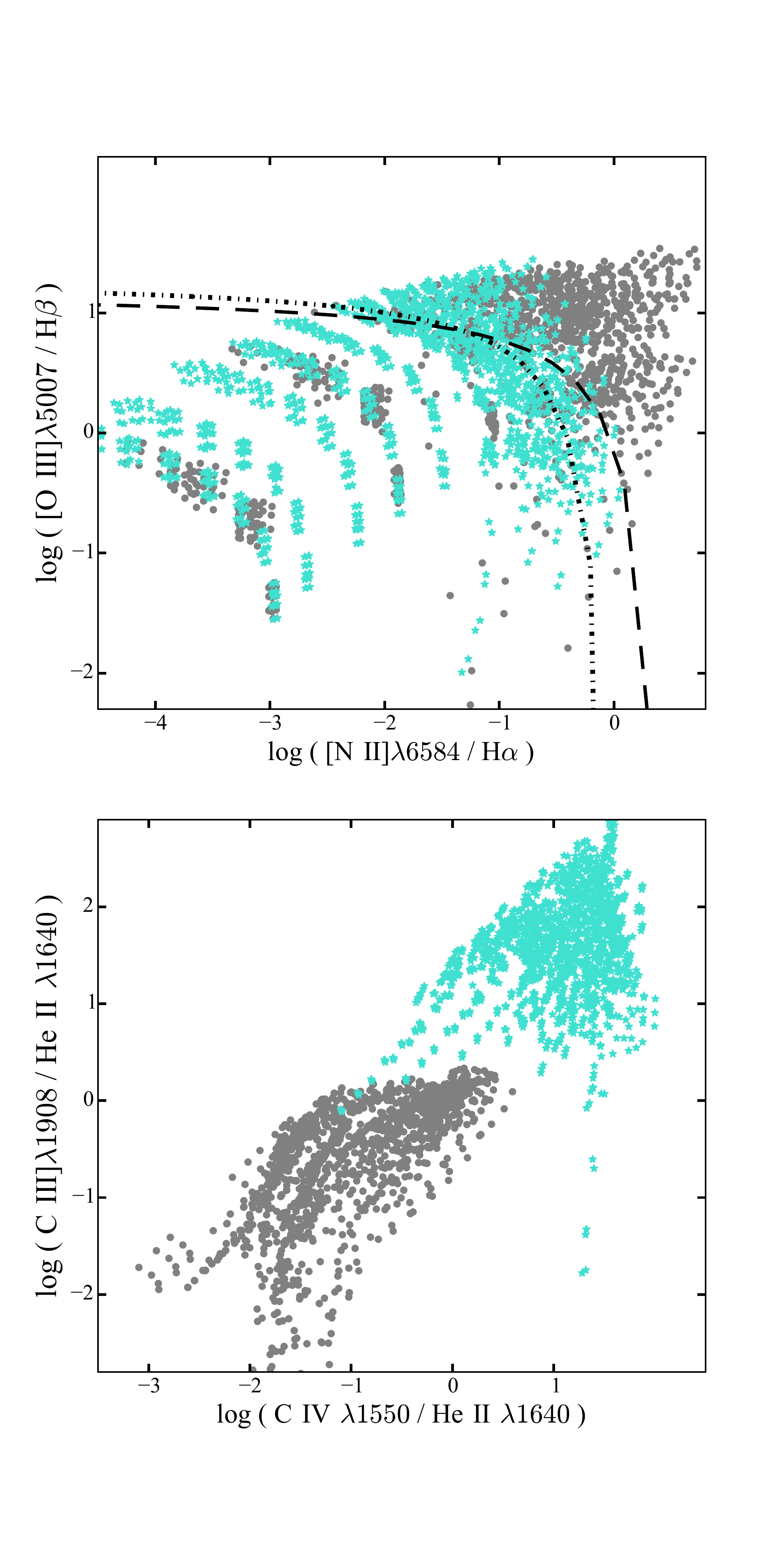} 
\caption{Distribution of AGN (gray dots) and SF (turquoise stars) models spanning full ranges in all the adjustable parameters listed in Table~\ref{tab:parameters} in the optical \oiii/H$\beta$ versus \nii/H$\alpha$ (BTP; top panel) and ultraviolet \ciiit/\heii\ versus \civt/\heii\ (bottom panel) diagnostic diagrams. In the top panel, the long-dashed and dot-dashed black lines indicate, respectively, the criteria of \protect\cite{kewley01} and \protect\cite{kauffmann03} to distinguish AGN from star-forming galaxies.}
\label{fig:BPT_UV_sep}
\end{center}
\end{figure}

The main outcome of the previous subsections is that models of nebular emission from active and inactive galaxies populate different areas of  diagrams defined by the luminosity ratios of several ultraviolet emission lines. So far, we have compared in a same diagram AGN and SF models in reduced ranges of gas density $n_{\rm H}$ and dust-to-metal ratio $\xi_{\rm d}$, for grids of ionization parameter $U_{\rm S}$, metallicity $Z$ and spectral power-law index $\alpha$. To fully assess how different emission-line ratios can help discriminate between nuclear activity and star formation, we must plot in a same diagram models spanning full ranges in all the adjustable parameters listed in Table~\ref{tab:parameters}.

As an example, we show in Fig.~\ref{fig:BPT_UV_sep} the full grids of AGN (gray dots) and SF (turquoise stars) models in the optical BPT (top) and ultraviolet \ciiit/\heii\ versus \civt/\heii\ (bottom) diagnostic diagrams. AGN and SF models can occupy similar regions of the BPT diagram. This similarity was already noted before for low-metallicity AGN and SF models (Fig.~\ref{fig:BPT1} of Section~\ref{sec:BPT}). SF models lying above the \citet{kewley01} separation criteria in Fig.~\ref{fig:BPT_UV_sep} (top) are those combining high ionization parameter ($\log U_{\rm S}\gtrsim-2$) and metallicity ($Z\gtrsim0.008$), a combination not commonly observed in the local Universe \citep[e.g.,][]{brinchmann04}. In contrast, Fig.~\ref{fig:BPT_UV_sep} (bottom) shows that the separation between AGN and SF models is far more distinct in the \ciiit/\heii\ versus \civt/\heii\ diagram. Galaxies with observed line ratios falling in the region of the diagram populated by AGN models will therefore be likely to be powered by nuclear activity.

In Appendix~\ref{app:sep}, we show the analogs of Fig.~\ref{fig:BPT_UV_sep} for the different ultraviolet emission-line ratios proposed to discriminate between active and inactive galaxies in Sections~\ref{sec:CHeC}--\ref{sec:Ne-based}, and which involve emission lines currently detected in observations of high-redshift galaxies \citep[][]{erb10,stark14}, i.e., \nv, \civt, \heii, \oiiit\ (and the corresponding individual doublet components), \niii, \siliiit\ (and the corresponding individual doublet components) and \ciiit. Figs~\ref{fig:sep1}, \ref{fig:sep2} and \ref{fig:sep3} show such diagnostic diagrams with \ciiit/\heii, \civt/\heii\ and \civt/\ciiit, respectively, on the ordinate axis. As seen previously, in some of these diagrams, SF models with highest metallicity ($Z\gtrsim0.03$), lowest density ($n_{\rm H}=10^2$~cm$^{-3}$) and lowest ionization parameter ($\log U_{\rm S}\lesssim-3$) can have line luminosity ratios similar to those of AGN models. This is the case for the \ciiit/\heii\ versus \nv/\civt, \oiii/\heii, \niii/\heii\ and \siliiit/\heii\ diagnostic diagrams (Fig.~\ref{fig:sep1}). Also, SF models with lowest metallicity ($Z=0.0001$) can overlap with AGN models in the \civt/\heii\ versus \nv/\heii, \nv/\civt, \nv/\niii, \niii/\heii\ and \siliiit/\heii\ diagrams (Fig.~\ref{fig:sep2}). Despite these limitations, over most of the parameter space, the diagrams in Figs~\ref{fig:sep1} and \ref{fig:sep2} remain good diagnostics of photoionization by nuclear activity versus star formation in galaxies. The overlap between AGN and SF models is larger in \civt/\ciiit\ versus \nv-based diagrams (top panels of Fig.~\ref{fig:sep3}), making these the least reliable diagnostics of active versus inactive galaxies investigated here.

\section{Summary and conclusions}\label{sec:conclusions}

We have developed a simple model to compute the nebular emission from narrow-line emitting regions of active galaxies in the presence of dust. These AGN models are parametrized in terms of the hydrogen density, ionization parameter, metallicity and dust-to-metal mass ratio of the ionized gas and the power-law index of the ionizing spectrum. We have shown that the models reproduce well the observed properties of active galaxies in standard diagrams defined by optical lines ratios, such as \oiii/H$\beta$, \nii/H$\alpha$, \siit/H$\alpha$ and \oi/H$\alpha$, used to discriminate between photoionization by nuclear activity and star formation in galaxies. In the same diagrams, the models of nebular emission from star-forming galaxies (SF models) recently proposed by G15 reproduce well the observed properties of SDSS galaxies. Given these premises, our main goal in the present study has been to extend the comparison between the properties of AGN and SF models to ultraviolet wavelengths and identify new emission line-ratio diagnostics of nuclear activity versus star formation. Such diagnostics should be particularly useful to interpret the rest-frame ultraviolet emission from high-redshift galaxies. 

We find that several combinations of the \heii\ recombination line with a collisionally excited metal line can serve, taken individually, as a good diagnostics of nuclear activity versus star formation in galaxies. This is the case for the luminosity ratios of \civt, \oiiit, \niii, \siliiit\ and \ciiit\ to \heii. The more numerous the available measurements of these different metal lines in addition to \heii, the more stringent the constraints on the nature of the ionizing source. Other ultraviolet emission-line ratios, such as \civt/\ciiit, \nv/\heii\ and \nv/\civt, do not individually allow a clear discrimination between active and inactive galaxies. However, information on these ratios will be useful to further investigate the physical parameters of the ionized gas, such as ionization parameter and metallicity, when combined with the luminosity ratio of either \civt, \oiiit, \niii, \siliiit\ or \ciiit\ to \heii\ (Figs~\ref{fig:UV_Villar} and \ref{fig:UV_NV}). We note that, for some combinations of gas density and dust-to-metal mass ratio at either extremely low ($Z\sim0.0001$) or high ($Z\gtrsim0.03$, i.e. greater than about twice solar) metallicity, AGN and SF models can occupy similar regions in some of these line-ratio diagnostic diagrams (see Section~\ref{sec:UV} for detail).

It is also interesting to note that AGN and SF models separate particularly well in various ultraviolet diagnostic diagrams involving Ne-based emission lines, such as \neiv, \neiii\ and \nev. In such diagrams, the luminosity ratios predicted by AGN and SF models can differ by up to 3 or 4 orders of magnitude. While this reflects the extreme faintness, and hence, unlikely observability, of some emission lines involved in the definition of these ratios, simple lower and upper limit on these emission-line ratios can already provide valuable clues about the nature of the ionizing source. For example, a detection of \neiv\ or \nev\ is likely to be associated to the presence of an active nucleus and, combined with other ultraviolet line, may also provide valuable information about the physical conditions in the ionized gas.

The AGN and SF models investigated in this paper compare well with currently available observations of rest-frame ultraviolet emission lines of active and inactive galaxies. In particular, our AGN models can account for the \nv, \civt, \heii\ and \ciiit\ emission-line properties of active galaxies at various redshifts from the sample of \citet[][and references therein]{dors14}. We find that the observed N-based emission-line ratios from this sample are best accounted for by models with slightly different parameters than those favored by other line ratios. This could be related to the traditional finding that AGN models around solar metallicity predict lower \nv\ line luminosity than observed in broad- and narrow-line regions of AGN \citep[e.g.,][]{osmer76,kraemer98}. This could also arise from a possible contamination of the AGN line-flux measurements by star formation (Section~\ref{sec:UVdata}). The SF models developed by G15, for their part, are in remarkable agreement with available measurements of (or upper limits on) the \nv, \civt, \heii, \oiiit, \niii, \siliiit\ and \ciiit\ emission-line properties of gravitationally lensed, low-mass star-forming galaxies at redshifts $z\sim2$ from \citet{stark14}. These ultraviolet emission lines are now detected routinely in quasars and galaxies out to high redshifts, although the lack of published measurements prevented us from inserting them in the present study \citep[e.g.,][]{humphrey07,hainline11,paris12,alexandroff13,paris14,lusso15}. Soon, high-quality ultraviolet spectroscopic observations will also be collected for large samples of galaxies out to the reionization epoch with future facilities such as \textit{JWST} and ground-based ELT. Our AGN and SF models will allow one to extract valuable information from these observations about the nature of the dominant ionizing sources in the early Universe. 

In this paper, we have not investigated the spectral signatures of photoionization by shocks, which can also contribute to the observed emission-line spectra of galaxies \citep{rich11}. The contribution by shocks to photoionization in the early Universe has been the subject of several studies. \citet{dopita11} argues that this could be substantial, while \citet{miniati04} and \citet{wyithe11} favor a negligible contribution. Unfortunately, existing shock models \citep[e.g.,][]{allen98,allen08} cannot be straightforwardly compared with the AGN and SF models considered here, because of the different ways in which the photoionized gas is parametrized. \cite{allen98,allen08} investigate ultraviolet emission-line diagnostics to discriminate between photoionization by shocks and nuclear activity. They propose the \cii/\ciiit\ versus \civt/\ciiit\ diagram as one of the most suitable discriminant between AGN and shocks with velocities less than $400\,\rm{km\,s}^{-1}$, while higher-velocity shocks overlap with AGN models in this diagram. We have used the AGN and SF models described in Section~\ref{sec:models} to investigate whether the \cii/\ciiit\ versus \civt/\ciiit\ diagram could also be useful to distinguish AGN from star-forming galaxies. In reality, we find that the 2 types of models overlap significantly in this diagram. We plan to investigate in a self-consistent way the ultraviolet spectral signatures of shocks versus AGN and SF models in a future study.

The ultraviolet line-ratio diagrams presented in this paper to discriminate between active and inactive galaxies are designed to help interpret current and future observations of the nebular emission from galaxies at all redshifts. These diagrams are also valuable to constrain the physical properties of photoionized gas, such as hydrogen density, ionization parameter, metallicity and dust-to-metal mass ratio. We note that, in this context, spatially resolved spectroscopy will be particularly useful to trace the different spatial scales of nuclear activity and star formation. To optimise the usefulness of the diagnostic diagrams presented in this paper for the interpretation of observed emission-line properties of galaxies, we are exploring statistical machine learning techniques to build an automated classifier able to probabilistically separate active from inactive galaxies (Stenning et al., in preparation). The spectral features most influencing this classification will also reveal which emission-line ratios might be considered as the most reliable discriminants between nuclear activity and star formation. Finally, it is worth mentioning that the AGN and SF models of nebular emission analysed in this paper are being incorporated in the Bayesian spectral interpretation tool BANGS (Chevallard et al., in preparation) to produce statistical constraints on the shape of the ionizing radiation and interstellar gas parameters.

\section*{Acknowledgements}

We acknowledge support from the ERC via an Advanced Grant under grant agreement no. 321323-NEOGAL. 

Funding for the creation and distribution of the SDSS Archive has been provided by the Alfred P. Sloan Foundation, the Participating Institutions,  the  National  Aeronautics  and  Space  Administration, the National  Science  Foundation, the US Department of Energy, the Japanese Monbukagakusho and the Max Planck Society. The SDSS website is http://www.sdss.org/.

Part of the analysis of this work makes use of the pyCloudy package developed by Christophe Morisset \citep{morisset13}. We thank Christophe Morisset for technical support and helpful comments. We also thank Stephanie Juneau for her assistance in selecting the SDSS observational sample. Finally, we thank the entire NEOGAL group for fruitful discussions and useful suggestions.

%%%%%%%%%%%%%%%%%%%%%%%%%%%%%%%%%%%%%%%%%%%%%%%%%%



%\bibliographystyle{mnras}


%%%%%%%%%%%%%%%%%%%%%%%%%%%%%%%%%%%%%%%%%%%%%%%%%%

%%%%%%%%%%%%%%%%% APPENDICES %%%%%%%%%%%%%%%%%%%%%

\appendix

\section{Ultraviolet line-ratio diagnostic diagrams of active and inactive galaxies}\label{app:sep}

\begin{figure*}
\begin{center}
\includegraphics[width=17.0cm]{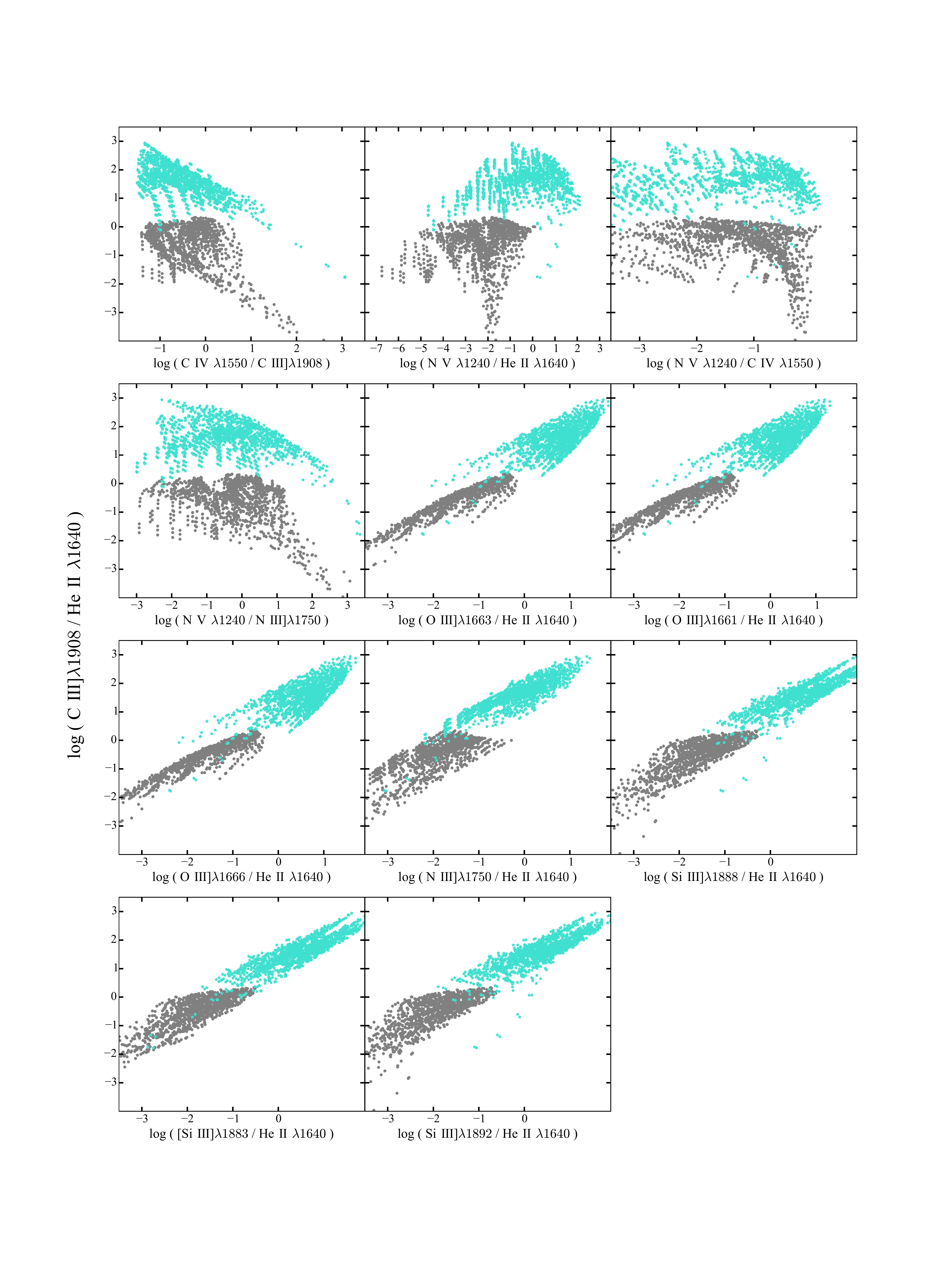} 
\caption{Distribution of AGN (gray dots) and SF (turquoise stars) models spanning full ranges in all the adjustable parameters listed in Table~\ref{tab:parameters} in ultraviolet line-ratio diagrams defined by \ciiit/\heii\ as a function of \civt/\ciiit, \nv/\heii, \nv/\civt, \nv/\niii, \oiiit/\heii, \oiiisf$\lambda1661$/\heii, \oiiisf$\lambda1666$/\heii, \niii/\heii, \siliiit/\heii, \siliii$\lambda1883$/\heii\ and \siliii$\lambda1892$/\heii.}
\label{fig:sep1}
\end{center}
\end{figure*}

\begin{figure*}
\begin{center}
\includegraphics[width=17.0cm]{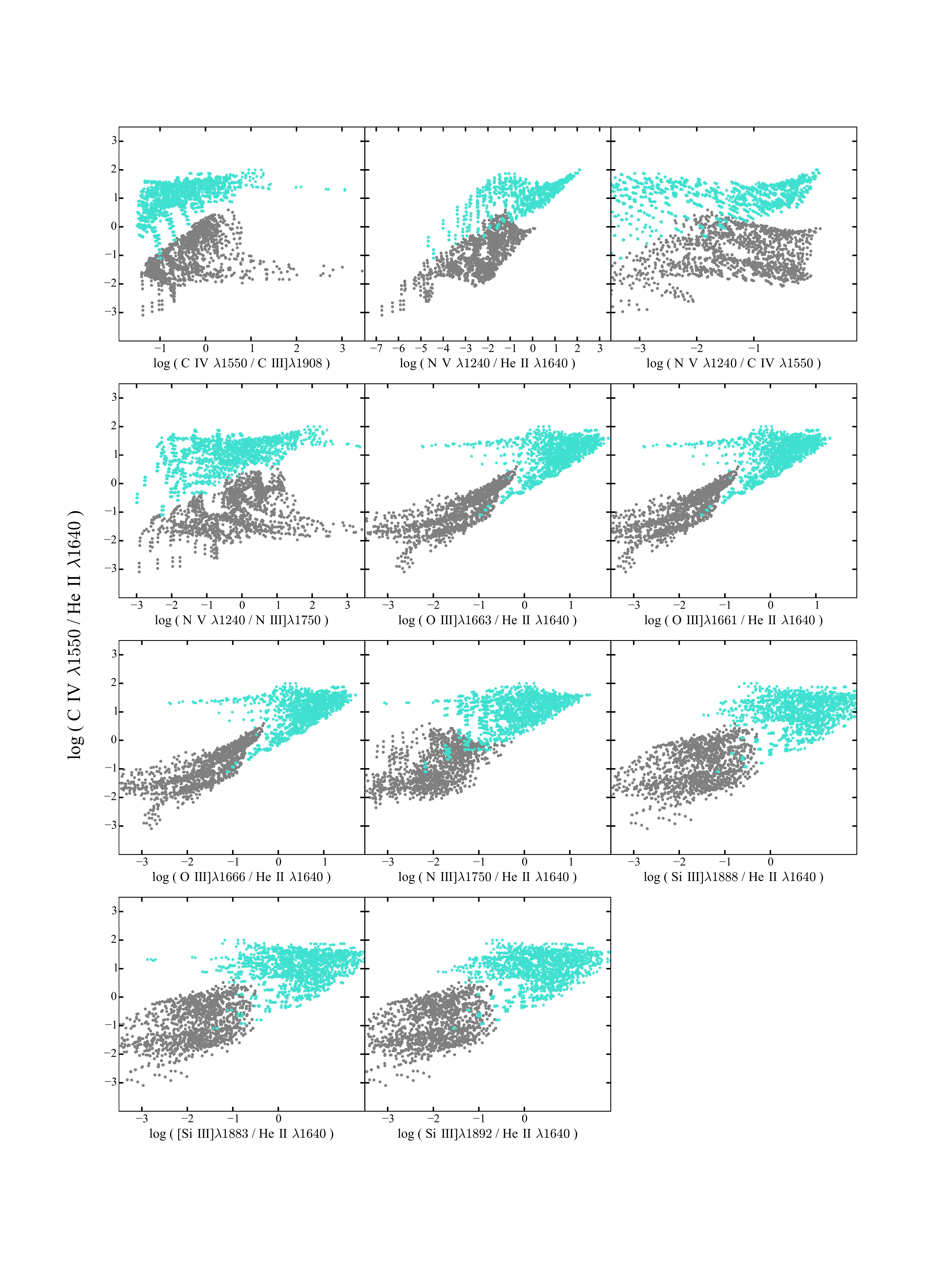} 
\caption{Distribution of AGN (gray dots) and SF (turquoise stars) models spanning full ranges in all the adjustable parameters listed in Table~\ref{tab:parameters} in ultraviolet line-ratio diagrams defined by \civt/\heii\ as a function of \civt/\ciiit, \nv/\heii, \nv/\civt, \nv/\niii, \oiiit/\heii, \oiiisf$\lambda1661$/\heii, \oiiisf$\lambda1666$/\heii, \niii/\heii, \siliiit/\heii, \siliii$\lambda1883$/\heii\ and \siliii$\lambda1892$/\heii.}
\label{fig:sep2}
\end{center}
\end{figure*}

\begin{figure*}
\begin{center}
\includegraphics[width=17.0cm]{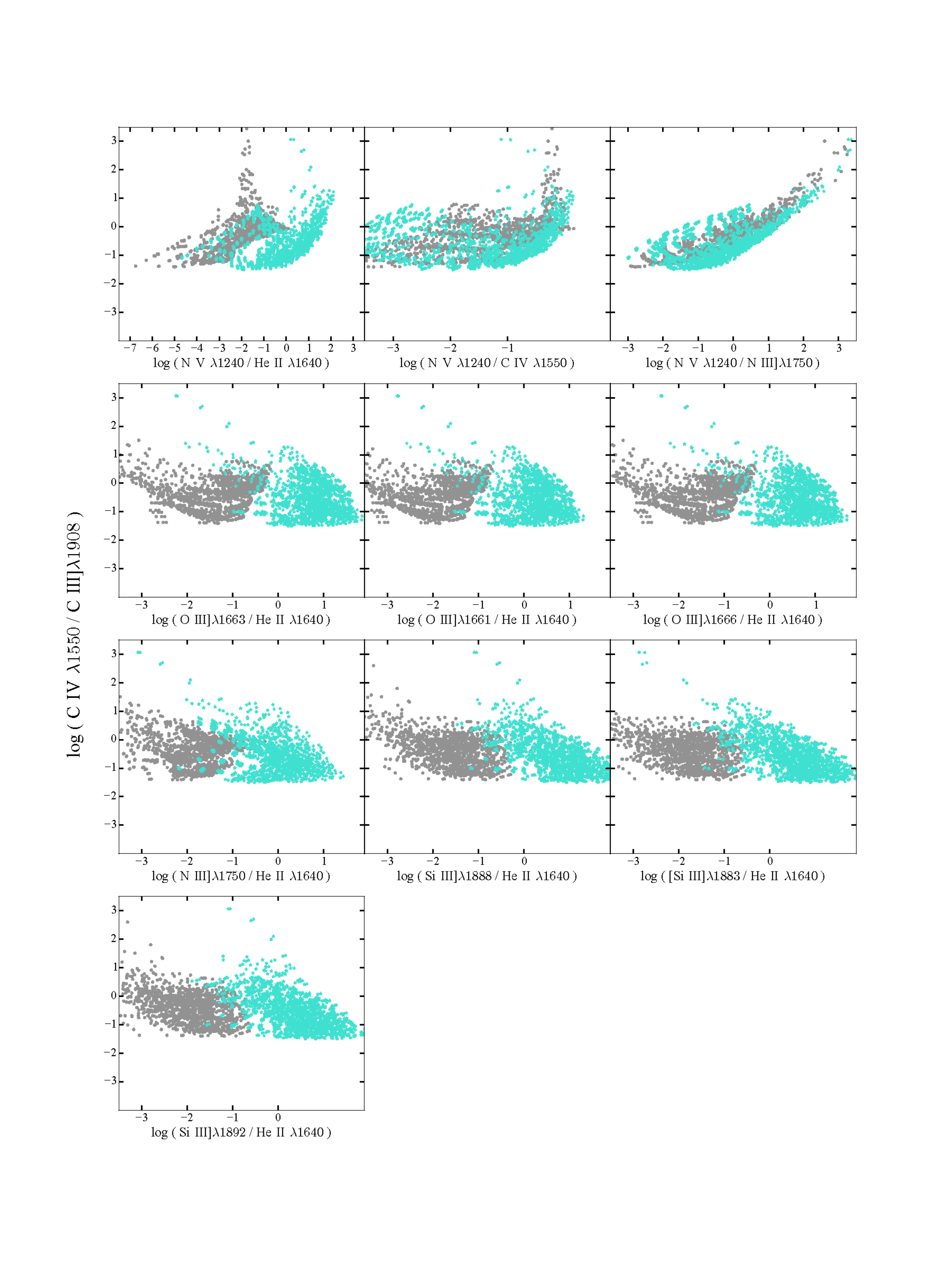} 
\caption{Distribution of AGN (gray dots) and SF (turquoise stars) models spanning full ranges in all the adjustable parameters listed in Table~\ref{tab:parameters} in ultraviolet line-ratio diagrams defined by \civt/\ciiit\ as a function of \nv/\heii, \nv/\civt, \nv/\niii, \oiiit/\heii, \oiiisf$\lambda1661$/\heii, \oiiisf$\lambda1666$/\heii, \niii/\heii, \siliiit/\heii, \siliii$\lambda1883$/\heii\ and \siliii$\lambda1892$/\heii.}
\label{fig:sep3}
\end{center}
\end{figure*}

Fig.~\ref{fig:sep1}, \ref{fig:sep2} and \ref{fig:sep3} shows diagnostic diagrams defined by the luminosity ratios of several ultraviolet emission lines currently detected in observations of high-redshift galaxies \citep[][]{erb10,stark14}, i.e., \nv, \civt, \heii, \oiiit\ (and the corresponding individual doublet components), \niii, \siliiit\ (and the corresponding individual doublet components) and \ciiit. In each panel of each figure, we show AGN (gray dots) and SF (turquoise stars) models spanning full ranges in all the adjustable parameters listed in Table~\ref{tab:parameters}. Figs~\ref{fig:sep1}, \ref{fig:sep2} and \ref{fig:sep3} show diagnostic diagrams with \ciiit/\heii, \civt/\heii\ and \civt/\ciiit, respectively, on the ordinate axis. We refer to Section~\ref{sec:separation} for more details about the usefulness of these diagrams to discriminate between nuclear activity and star formation in galaxies.

\bsp	
\label{lastpage}
\end{document}